\documentclass[%
 aip,
 amsmath,amssymb,
preprint,%
]{revtex4-1}

\usepackage{graphicx}
\usepackage{dcolumn}
\usepackage{bm}
\usepackage{subcaption}
\usepackage{verbatim}

\usepackage[utf8]{inputenc}
\usepackage[T1]{fontenc}
\usepackage{mathptmx}
\usepackage{etoolbox}
\usepackage{calrsfs}
\usepackage{xcolor}
\DeclareMathAlphabet{\pazocal}{OMS}{zplm}{m}{n}

\makeatletter
\def\@email#1#2{%
 \endgroup
 \patchcmd{\titleblock@produce}
  {\frontmatter@RRAPformat}
  {\frontmatter@RRAPformat{\produce@RRAP{*#1\href{mailto:#2}{#2}}}\frontmatter@RRAPformat}
  {}{}
}%
\makeatother
\begin{document}

\preprint{AIP/123-QED}

\title{Effect of sweep angle on three-dimensional vortex dynamics over plunging wings}
\author{Alex Cavanagh}
\affiliation{Autonomous Systems and Connectivity, School of Engineering, University of Glasgow, G12 8QQ, Glasgow, United Kingdom
}%
\author{Chandan Bose}%
\affiliation{ 
Aerospace Engineering, School of Metallurgy and Materials, University of Birmingham, B15 2TT, Birmingham, UK
}%
\author{Kiran Ramesh}
\affiliation{Autonomous Systems and Connectivity, School of Engineering, University of Glasgow, G12 8QQ, Glasgow, United Kingdom
}%
\email{kiran.ramesh@glasgow.ac.uk}

\date{\today}

\begin{abstract}
The effects of sweep angle and reduced frequency on the leading-edge vortex (LEV) structure over flapping swept wings in the Reynolds number ($Re$) range of $\mathbf{O}(10^4)$ are yet to be completely understood. With increasing interest in designing bio-inspired micro-air-vehicles (MAVs), understanding LEV dynamics in such scenarios is imperative. This study investigates the effects of three different sweep angles ($\Lambda = 0^\circ$, $30^\circ$ and $60^\circ$) on LEV dynamics through high-fidelity improved delayed detached eddy simulation (IDDES) to analyze the underlying flow physics. Plunge ramp kinematics at two different reduced frequencies ($k = 0.05$ and $0.4$) are studied to investigate the unsteady motion effects on LEV characteristics. The leading-edge suction parameter (LESP) concept is applied to determine LEV initiation, and the results are verified against flow field visualization for swept-wing geometries. The force partitioning method (FPM) is used to investigate the spanwise lift distribution resulting from the LEV. Distinct peaks in the lift coefficient occur for the high reduced frequency case due to the impulse-like plunging acceleration. This causes the LEV to detach from the leading edge more quickly and convect faster, significantly affecting the lift generated by the wing. As reduced frequency increases, the LEV breakdown mechanism switches from vortex bursting to LEV leg-induced instabilities. These results provide insights into the complex vortex structures surrounding swept wings at $Re = 20,000$, and the impact both sweep angle and reduced frequency have on the lift contribution of these flow features.
\end{abstract}

\maketitle

\section{Introduction}

It has been long known that the flapping motions of natural fliers utilize unsteady aerodynamics to enhance lift; however, the mechanisms by which this lift increment is achieved were not understood initially. Ellington \textit{et al.} \cite{ellington_1996} first discovered the leading-edge vortex (LEV) as the answer. The shedding of an LEV is often associated with dynamic stall, characterized by a delay in stall onset considerably beyond the static stall limit \cite{mccroskey_1982}. The authors found that, in insect flight, vortex stability is aided by LEV convection in the spanwise direction towards the tip. 

The primary LEV is formed through the roll-up of the shear layer at the leading edge of a wing, sheds, and convects downstream. As the LEV grows in strength and re-attaches to the wing surface, flow is induced towards the leading edge between the LEV and wing, causing secondary separation. This leads to a counter-rotating secondary vortex under the primary vortex \cite{anderson_1991}. LEV formation also occurs in numerous other lifting wing applications, such as over helicopter rotor blades.

The LEV was first recognized as detrimental to helicopter rotor blade performance \cite{mccroskey_1981}. Hence, research was carried out to prevent its formation. Inspired by insect flight, recent studies have considered the low Reynolds numbers and high reduced frequencies typical of biological flight \cite{eldredge&jones_2019}. Pitching kinematics were most commonly analyzed \cite{granlund_2013, stevens&babinsky_2017, yilmaz&rockwell_2012, eldredge_2009, jantzen_2014, visbal_2017, green_2011, hord&lian_2016, yilmaz_2010}, however some studies have also considered plunging motions \cite{calderon_2013, calderon_2013a, calderon_2014, fishman_2017, visbal_2013, yilmaz&rockwell_2010}, rotation \cite{beals&jones_2015,carr_2013,carr_2015,medina&jones_2016,ozen&rockwell_2012,venkata&jones_2013}, translation \cite{devoria&mohseni_2017,mancini_2015,mulleners_2017}, and their combinations \cite{ol_2009, manar_2016, percin&vanoudheusden_2015}. Bio-inspired micro air vehicle (MAV) design aims to utilize the beneficial unsteady characteristics of natural flight \cite{chellapurath_2020}. To that end, engineers attempt to stabilize the LEV near the wing surface to benefit from its improved transient lift. 

Biological fliers also often enhance flight efficiency through their wing shape. They utilize various wing planforms to enable agile flight dynamics. Swept wings are ubiquitous in nature, from the common swift (\textit{Apus apus}) to the hawk moth (\textit{Manduca sexta}). Through changes to its wing sweep and aspect ratio depending on the flight regime, the common swift's aerodynamic performance is seen to be significantly improved \cite{lentink_2007, weiss_2003, rayner_1988, azuma_2006}. Previous studies on simplified wing geometries undergoing unsteady kinematic motions have revealed the governing parameters, underpinning the LEV-induced lift enhancement. Plunging high aspect-ratio wings with different sweep angles were experimentally studied by Chiereghin \textit{et al.} \cite{chiereghin_2017} at $Re = 20,000$. The influence of the LEV on forces and moments was found to increase with higher sweep angles. In a subsequent study, Chiereghin \textit{et al.} \cite{chiereghin_2020} compared results from an unswept wing to a $40^\circ$ swept wing to analyze sweep angle effects on the three-dimensional (3D) nature of the LEV. Son \textit{et al.} \cite{son_2022} numerically extended this work up to a maximum reduced frequency of $k = 3$, to understand the previously measured spanwise instabilities and deformation of the LEV \cite{chiereghin_2017}. However, Son \textit{et al.} only studied an unswept wing. Hence, the extent to which these conclusions apply to swept wings remains unexplored. Computational fluid dynamics (CFD) simulations are an alternative approach to experimentation that allows a more significant number of cases to be analyzed due to the lower monetary cost. However, the choice of modelling approach requires attention. 

For example, complex flow features such as vortex interactions and breakdown remain unclear in Reynolds-averaged Navier-Stokes (RANS) simulations due to the time-averaging approach. Therefore, recent studies have opted higher-order simulation approaches to obtain a detailed overview of the flow-field structures, incuclide LEV, surrounding the wings \cite{hammer_2023, garmann&visbal_2022}. Visbal and Garmann \cite{visbal&garmann_2019} conducted large eddy simulations (LES) of pitching wings at $Re = 2 \times 10^5$ and $k \approx 0.2$ to investigate sweep angle effects on dynamic stall. Arch vortices were formed about the wing apex and located further outboard as the sweep angle increased. The swept geometry causes dynamic tip stall, resulting in the loss of suction upstream of the displaced tip vortex. The vortex leg interacts strongly with the tip flow at a higher sweep angle, and a sudden tip stall occurs. Although the authors provided useful qualitative insights into the LEV flow field, the quantitative effects of the LEV on the wing lift distribution were not studied in detail.

Recently, to quantify the effect of vorticity on the aerodynamic loading of lifting surfaces, the force and moment partitioning method (FMPM) is proposed by Zhang \textit{et al.} \cite{zhang_2015}. This allows the contribution of individual vortex structures underpinning the force generation mechanisms to be investigated. Similar analyses have been carried out in past studies using impulse theory \cite{graham_2017} and the vortex force map method \cite{li&wu_2018}. Zhu and Breuer \cite{zhu&breuer_2023} applied the FMPM to pitching swept wings and reported that wing sweep had a small effect on force generation. It has been shown that the sweep angle influence on moment distribution increases due to a larger effective moment and becomes asymmetric as sweep increases due to the variation in the effective pitch axis along the span \cite{zhu_2021}. This results in non-intuitive distributions of the moment, such as the LEV contributing negatively to the moment towards the tip as the sweep angle increases. Menon \textit{et al.} \cite{menon_2022} studied the contributions of spanwise and cross-span vortices to lift generation using the FMPM. Swept wings at fixed angles of attack \cite{zhang&taira_2022} and delta wings \cite{li_2020} have also been studied using the FMPM.

Developing low-order models (LOMs) allows the key flow physics that govern LEV dynamics to be uncovered. One approach that determines the onset of LEV initiation is the leading-edge suction parameter (LESP), proposed by Ramesh \textit{et al.} \cite{ramesh_2014}. 
LESP is equivalent to the circulation at the leading edge, which depends on the airfoil's downwash. The authors discovered a constant critical value of LESP ($LESP_{crit}$) under kinematic changes for any airfoil at a fixed Reynolds number. When $LESP_{crit}$ is exceeded, a vortex of strength relative to the excess LESP is shed at the leading edge to limit LESP to the critical value. For finite wings, LEV dynamics are significantly more complicated due to spanwise velocities and pressure gradients, rotational effects, and vortex interactions \cite{maxworthy_2007, harbig_2014, wojcik&buchholz_2014, wong&rival_2015, limacher_2016}. Therefore, the development of LOMs has primarily focussed on 2D airfoils. However, LESP has been used to determine LEV initiation on a finite wing using an unsteady vortex lattice method (UVLM) \cite{hirato_2019, hirato_2021}. This required an empirically determined scaling factor to match the values of Ramesh \textit{et al.} \cite{ramesh_2014} due to the assumption that other terms are zero over the length of the first UVLM panel \cite{aggarwal_2013}. Hence, there is a significant scope to determine the applicability of the LESP concept, originally intended for 2D cases, to swept wings. This is taken up in this study.

This work extends upon existing literature by exploring the effects of sweep on LEV dynamics in the context of plunging wings in the $Re = \mathbf{O}(10^4)$ regime. We aim to answer to what extent sweep angle and unsteady plunging kinematics affect LEV initiation and structure and how this affects the wing lift distribution. We use high-fidelity improved delayed detached eddy simulation (IDDES) to capture the intricacies of the separated vortex structures using LES while modeling the laminar boundary layer with RANS. We verify the use of LESP for swept wings using a method based on the shear layer velocity that requires no empirical correction. Finally, the force partitioning method (FPM) is used to provide quantitative data for the vorticity-induced lift distribution so that the effects of sweep and kinematics can be analyzed.

An outline of the methodology used to calculate LESP is presented in section II.A, with the FPM overview provided in section II.B. The computational methodology is provided in section III, with the geometries and kinematics described in section III.A, followed by an overview of the detached eddy simulation method in section III.B. Validation of numerical results through comparison to experimental results and mesh verification is provided in section III.C. The effects of sweep angle and reduced frequency are discussed regarding lift coefficient in section IV.A, followed by the discussion on the LEV initiation in section IV.B, vortex structures in section IV.C, LEV convection rate in section IV.D, and LEV lift distribution in section IV.E. Finally, salient outcomes of the present study are concluded in section V.

\section{Theoretical Models}

\subsection{Leading-edge Suction Parameter}

In this work we use the LESP concept to study LEV formation over swept wings, using the method derived by Ramesh \cite{ramesh_2020} that uses matched asymptotic expansions to derive an expression for the velocity at the leading edge dependent only on LESP, as shown in equation \ref{vel_LE_A0}.

  \begin{equation}\label{vel_LE_A0}
    u_{LE} = \sqrt{\frac{2}{r_{LE}}}UA_0.
  \end{equation}

 Here $A_0 = \pazocal{L}$ and $r_{LE}$ represents the radius of the airfoil at the leading edge, where $\pazocal{L}$ represents LESP. This overcomes the limitations of relying on airfoil discretization in a vortex lattice-based method by using unsteady thin-airfoil theory (UTAT). The method has been extended by Martínez \textit{et al.} \cite{martinez_2022} by relating the velocity at the shear layer edges to LESP by assuming that the LEV is the dominant flow feature. This simplifies the literature \cite{fage&johansen_1927} expression in equation \ref{shear_layer} to equation \ref{vel_LE}.

  \begin{equation}\label{shear_layer}
    \frac{d\Gamma_s}{dt} = \frac{1}{2}(V_1^2-V_2^2),
  \end{equation}

  \begin{equation}\label{vel_LE}
  \Dot{\Gamma}_{LE} = \frac{1}{2}(u_{LE}^2),
  \end{equation}

where $\Gamma_s$ is shear layer circulation, $V_1$ and $V_2$ are shear layer edge velocities and $\Dot{\Gamma}_{LE}$ is the rate of change of leading edge circulation. We extend this method by applying the 2D theory to a 3D case by considering cross-sectional planes parallel to the root plane at various spanwise locations. This method is thus a 2.5D method.

\subsection{Force Partitioning Method}

We use the force partitioning method (FPM) developed by Menon and Mittal \cite{menon&mittal_2021,menon&mittal_2021b} to quantify the vorticity-induced forces on the wings. The FPM consists of projecting the Navier-Stokes momentum equation onto the gradient of an influence potential field to decompose the force on the wing into its constituent components, the major components being force due to unsteady effects, such as transient kinematics, viscosity and vorticity. The influence potential $\phi$ is a function of the wing geometry \textit{B} and position only and satisfies a Laplace equation with Neumann boundary conditions, as shown in equation \ref{influence_potential}.

  \begin{equation}\label{influence_potential}
  \nabla^2\phi_i = 0, \; \text{with} \: \textbf{n}\cdot\nabla\phi_i = 
    \begin{cases}
        n_i & \text{on} \: B\\
        0 & \text{on} \: \Sigma
    \end{cases}
    \: \text{for} \: i = 1,2,3.
  \end{equation}

Here, \textbf{n} is the normal vector, defined as negative pointing into the domain, where $n_i$ is the component of \textbf{n} in the \textit{i} Cartesian coordinate direction and $\Sigma$ is the chosen outer domain boundary for integration of forces. The influence potential describes the spatial influence of vorticity on the total force on the wing and by integrating over volumes of interest \textit{V} the total force due to vorticity $F_i^\omega$ within the specified volume $V_f$ can be determined as below.

  \begin{equation}\label{vorticity_force}
  F_i^\omega = -2\int_{V_f}{Q\phi_i} \: dV \; \text{for} \: i = 1,2,3.
  \end{equation}

$Q$ is the second invariant of the velocity gradient tensor, defined as $Q = \frac{1}{2}(||\boldsymbol{\Omega}||^2-||\textbf{S}||^2$), where $\boldsymbol{\Omega}$ represents rotation and $\textbf{S}$ represents strain \cite{jeong&hussain_1995}. Hence, regions of positive $Q$ are often used to detect vortices within the flow field.

\section{Computational Methodology}

\subsection{Problem Definition}

To investigate sweep angle effects, three wing geometries are considered in this study with sweep angles of $\Lambda = 0^\circ$, $30^\circ$ and $60^\circ$. The wing geometries have a constant aspect ratio of 3, which closely resembles that of biological fliers \cite{lentink&dickinson_2009} and a NACA0008 airfoil profile. The prescribed kinematic motion being investigated is a smoothed Eldredge ramp \cite{eldredge_2009, jantzen_2014}, representing a vertical gust, described in equation \ref{kinematics} with the smoothing parameter \textit{a} defined by Granlund \textit{et al.} \cite{granlund_2013} in equation \ref{smoothing}.

  \begin{equation}\label{kinematics}
  \frac{\Dot{h}}{c} = \frac{k}{a}\log\left[\frac{\cosh(a(t-t_1)}{\cosh(a(t-t_2))}\right]+ \frac{(\Dot{h}/c)_{amp}}{2},
  \end{equation}

    \begin{equation}\label{smoothing}
  a = \frac{\pi^2}{4(t_2-t_1)(1-\sigma)}.
  \end{equation}

\textit{k} is the chord-reduced frequency, $t_1$ and $t_2$ represent the start and end of the plunge ramp respectively and the chord-reduced plunge amplitude $(\Dot{h}/c)_{amp}$ is fixed as 0.25. The maximum plunge displacement is one chord length \textit{c}. The smoothing parameter $\sigma$ is equal to 0.8. The plunge rate $\Dot{h}$ corresponds to an effective pitch angle \cite{mcgowan_2011}, with the maximum plunge rate being equivalent to $\alpha \approx 14.5^\circ$. This was found to be sufficiently large enough to ensure that a coherent LEV formed for both reduced frequencies considered ($k = 0.05, 0.4$). The pitch angle is fixed at $0^\circ$. The set of 6 cases described in Table \ref{tab:cases} are considered in this study to investigate the effects of sweep angle and reduced frequency on LEV dynamics.

\begin{table}[h]
    \centering
    \caption{Case parameters}
    \begin{tabular}{c|c}
    \hline
    Parameter & Values \\
    \hline
    ${\Lambda}$ & $0^\circ, 30^\circ, 60^\circ$ \\
    \textit{k} & 0.05, 0.4 \\
    \end{tabular}
    \label{tab:cases}
\end{table}

\subsection{Detached Eddy Simulation}

DES is a two-part model that employs the conventional RANS approach for attached boundary layers, with separated large eddies simulated by LES. As complex 3D eddies are resolved directly, this makes DES more appropriate for highly-separated flows than RANS models, while requiring significantly less computational resources than LES. IDDES improves upon the original DES and delayed-DES (DDES) models by resolving the log-layer mismatch between modeled and resolved layers present in those models by adjusting the resolved log-layer based on the increase of resolved near-wall turbulent activity \cite{shur_2008}. The sub-grid length scale is redefined to also be dependent on the maximum grid spacing, as well as the distance to the wall. The IDDES governing equation for the transport of modified turbulence viscosity $\tilde{{\nu}}$ is given by equation \ref{IDDES}, with length scale $\tilde{d}$ given by equation \ref{d_tilde}, where $SMALL$ is used to prevent a singularity.

  \begin{eqnarray}\label{IDDES}
    \frac{D}{Dt}(\rho\tilde{\nu}) = &&\nabla \cdot (\rho D_{\tilde{\nu}}\tilde{\nu}) + \frac{C_{b2}}{\sigma_{\nu_t}} \rho |\nabla\tilde{\nu}|^2 + C_{b1} \rho \tilde{S} \tilde{\nu}(1 - f_{t2}) \nonumber\\&&- (C_{w1} f_w - \frac{C_{b1}}{\kappa^2} f_{t2}) \rho \frac{\tilde{\nu}^2}{\tilde{d}^2} + S_{\tilde{\nu}},
  \end{eqnarray}

  \begin{equation}\label{d_tilde}
    \tilde{d} = max(\tilde{f_d}(1 + f_e) L_{RAS} + (1 - \tilde{f_d})L_{LES}, SMALL).
  \end{equation}

\begin{figure}
\centering
\includegraphics{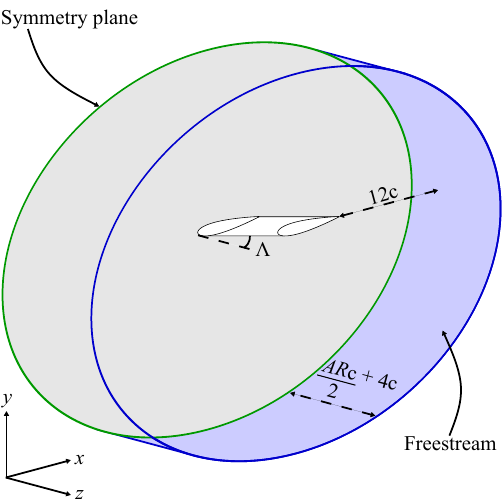}
\caption{Schematic diagram of the computational domain. The symmetry boundary condition is colored gray and the freestream boundary condition is colored blue. A wing of arbitrary sweep angle $\Lambda$ is depicted with zero pressure gradient boundary condition. The freestream flow is in the positive $x$ direction. The coordinate system origin is along the leading edge at the wing root. }
\label{fig:domain_schematic}
\end{figure}

The open-source CFD code library OpenFOAM is used to conduct simulations. The finite volume method is used to solve the time-dependent, incompressible Navier-Stokes equations. Discretization of the time derivatives is achieved using a second-order accurate backward scheme. Second-order accurate Gaussian integration schemes are also chosen for the gradient, divergence and Laplacian terms. The pressure implicit with splitting of operators (PISO) algorithm is used to achieve pressure-velocity coupling. The Spalart-Allmaras model is used for turbulence closure, as its effectiveness has been demonstrated for a wide range of unsteady separated external aerodynamics flows \cite{ramesh_2014,mcgowan_2011}. The no-slip boundary condition is applied to the airfoil surface and the far field is given the freestream (inlet/outlet) boundary condition. This acts as a zero-gradient condition when fluid is exiting the domain and as a fixed value condition, equal to the freestream value, otherwise. A symmetry plane is defined at the wing root plane to enable the simulation of half the symmetrical flow field. A schematic diagram of the computational domain is provided in Fig. \ref{fig:domain_schematic}.

A body-fitted mesh is constructed around the wing with an initial wall-normal cell spacing of $1.24 \times 10^{-4}c$ and extruded up to a distance of $12c$. Up to a distance of $2c$, the growth rate and maximum cell size are limited to achieve the cubic cell requirement of DES grids within the focus region, with a target grid spacing of $\Delta_0 = 0.015c$ chosen \cite{spalart_2001}. The O-mesh contains 249 chord-wise cells, with increased resolution towards the leading and trailing edges. The aspect ratio 3 wings to be simulated contains 250 spanwise cells, with reduced spacing towards the wingtip. The spanwise domain extends 4 chord lengths beyond the wingtip, with hyperbolic growth from the wingtip, with the first layer height being the same as in the normal direction to ensure $y^+ < 1$. The chord length is $c = 0.08$ m, freestream velocity $\textbf{U}_\infty = 0.16$ ms\textsuperscript{-1} and kinematic viscosity $\nu = 6.4 \times 10^{-6}$ m\textsuperscript{2}s\textsuperscript{-1} to give Reynolds number $Re = 2 \times 10^4$.

\subsection{Validation and Verification}

\begin{figure}
\centering
\includegraphics{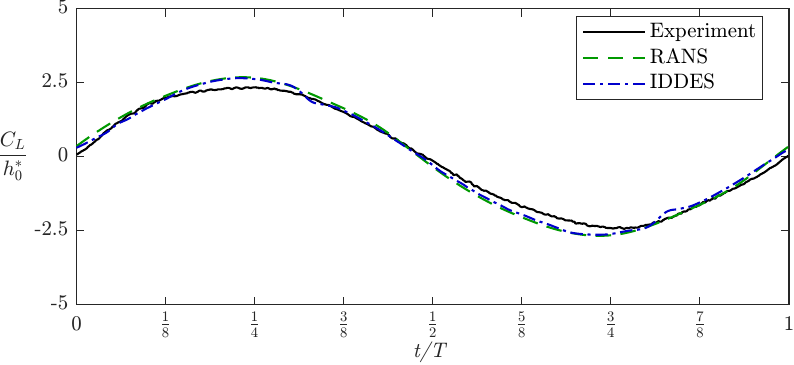}
\caption{Comparison between experimental results, RANS results and IDDES results for an unswept wing with $\textit{AR} = 3$ undergoing a sinusoidal plunging motion with a chord-reduced plunge amplitude ${h^*_0} = 0.5$.}
\label{fig:Cl_RANS_IDDES}
\end{figure}

\begin{figure}
\centering
\includegraphics{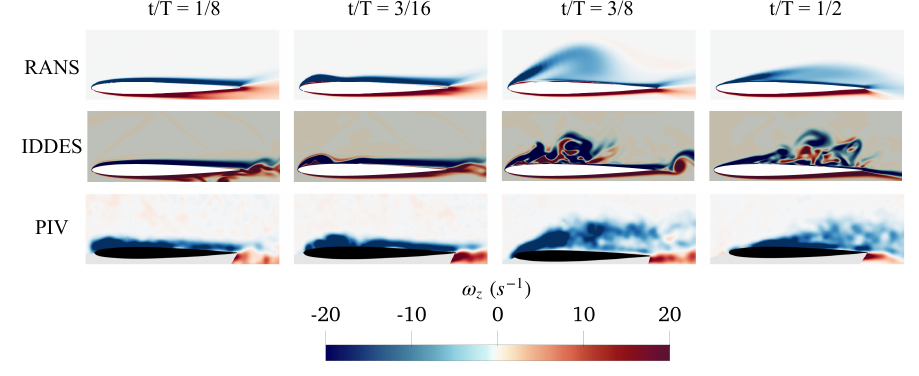}
\caption{Quarter-span spanwise vorticity distribution comparison between RANS results, IDDES results and experimental results for the plunging wing with aspect ratio 3.}
\label{fig:vorticity_validation}
\end{figure}

The mesh setup was validated against experiments and also compared against RANS simulation results for an aspect ratio three, $30^\circ$ swept wing case, undergoing a sinusoidal plunge motion at $Re = 10^4$, as in the work of Bird \textit{et al.} \cite{bird_2022}. The results of this validation are shown by the lift coefficient plot in Fig. \ref{fig:Cl_RANS_IDDES}. The RANS and IDDES results over-predict the peak experimental value by 14\% and 12\% respectively. There is a phase difference in global minimum between experimental and simulation results, with the IDDES result being closer to the experimental location. The RANS result shows a smooth, sinusoidal variation in lift coefficient throughout the motion, whereas the IDDES result has some deviations due to unsteady effects resulting from LEV breakdown, as demonstrated in Fig. \ref{fig:vorticity_validation} at $\textit{t/T} = 3/8$. The present results obtained from the IDDES simulations show close agreement with the reference experimental results. Hence, we use this approach for further simulations in this study.

Grid independence is verified by comparing the lift coefficient at the global minimum for each of three successive mesh refinement levels in Table \ref{tab:mesh_verification}. The percentage error to the medium mesh value is also reported. The mesh refinement was performed by successively refining the mesh size in the focus region around the wing, outside of the boundary layer mesh. Each mesh had a target grid spacing of $\Delta_0 = 0.018c$, $\Delta_0 = 0.015c$ and $\Delta_0 = 0.01c$ as refinement increased. The verification tests were performed for the moderately swept wing undergoing plunge velocity ramp kinematics at a reduced frequency of $k = 0.4$. The difference in the result between each mesh is small, although the relative error between the medium and fine meshes is an order of magnitude smaller than the error between the coarse and medium mesh. Therefore, the medium-fidelity mesh was selected to ensure solution accuracy while limiting computational expense.

\begin{table}[h]
    \centering
    \caption{Lift coefficient comparison between successive levels of mesh refinement at the global minimum.}
    \begin{tabular}{c|c|c}
    \hline
    Refinement level & $C_L$ & Percentage error (\%) \\
    \hline
    Coarse $(347 \times 219 \times 365)$ & -2.6511 & 0.26\\
    Medium $(504 \times 255 \times 365)$ & -2.6581 & --\\
    Fine $(567 \times 323 \times 365)$ & -2.6584 & -0.01\\
    \end{tabular}
    \label{tab:mesh_verification}
\end{table}

\section{Results and Discussion}

\subsection{Lift coefficient}

\begin{figure}
\centering
\begin{subfigure}{0.49\textwidth}
    \includegraphics{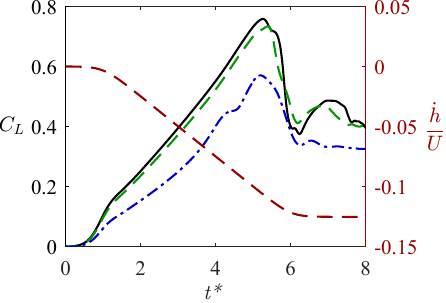}
    \caption{$k = 0.05$}
    \label{fig:C_L_first}
\end{subfigure}
\hfill
\begin{subfigure}{0.49\textwidth}
    \includegraphics{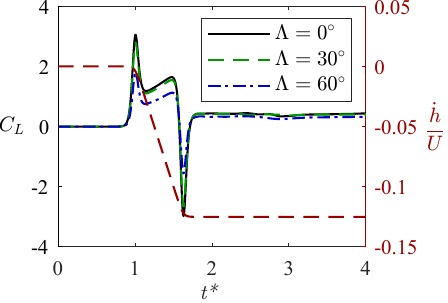}
    \caption{$k = 0.4$}
    \label{fig:C_L_second}
\end{subfigure}
  
\caption{Lift coefficient comparison for all swept wing geometries, with the prescribed plunge velocity kinematics shown on the right axis. The legend also applies to Figs. \ref{fig:LEV_3d} and \ref{fig:C_L_omega}.}
\label{fig:C_L}
\end{figure}

Fig. \ref{fig:C_L} shows the time histories of lift coefficient for each geometry at both the low (Fig. \ref{fig:C_L_first}) and high (Fig. \ref{fig:C_L_second}) reduced frequencies. This is overlaid with the plunge ramp kinematics to aid identification of the salient features. 

Fig. \ref{fig:C_L_first} shows a smooth variation of lift coefficient, with a nearly linear increase during the ramp-down motion. At the start and end of the ramp, opposing acceleration directions produce opposite effects on the lift coefficient. The unswept and $30^\circ$ swept wings have similar lift magnitudes throughout, with the $60^\circ$ swept wing having a significantly lower lift. At $t^* = 5.5$, there is a significant reduction in lift coefficient for all three wings, although this is more abrupt for the two lower sweep angle cases. This is due to the deceleration of the wing as the plunge rate becomes constant. 

Secondly, in Fig. \ref{fig:C_L_second}, unlike the low reduced frequency case, large added-mass spikes occur at the beginning and end of the plunge cycle due to the impulse-like acceleration of the wing at this time. During the ramp-down motion, the lift coefficient increases linearly for all three wings and there is a non-linear decrease in lift coefficient as the sweep angle increases at these times, similar to the low reduced frequency cases. However, the magnitude of the lift is approximately three times the magnitude. Once the plunge rate becomes steady, the lift coefficient for all three wings remains fairly constant and at a similar value of $C_L \approx 0.4$, which is comparable to the low reduced frequency cases as both motions are plunging at the same rate.

\subsection{LEV initiation}

\begin{figure}
\centering

\begin{subfigure}{0.32\textwidth}
    \includegraphics{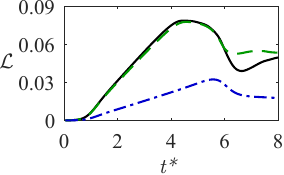}
    \caption{$k = 0.05$, $\Lambda = 0^\circ$}
    \label{fig:LESP_first}
\end{subfigure}
\hfill
\begin{subfigure}{0.32\textwidth}
    \includegraphics{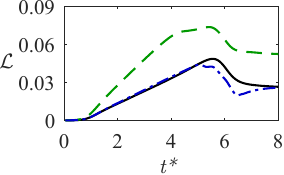}
    \caption{$k = 0.05$, $\Lambda = 30^\circ$}
    \label{fig:LESP_second}
\end{subfigure}
\hfill
\begin{subfigure}{0.32\textwidth}
    \includegraphics{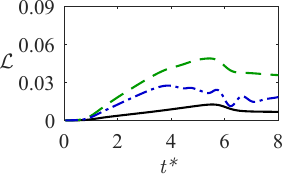}
    \caption{$k = 0.05$, $\Lambda = 60^\circ$}
    \label{fig:LESP_third}
\end{subfigure}

\hfill
\begin{subfigure}{0.32\textwidth}
    \includegraphics{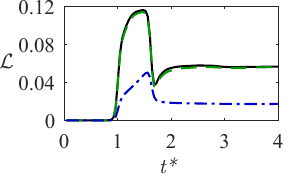}
    \caption{$k = 0.4$, $\Lambda = 0^\circ$}
    \label{fig:LESP_fourth}
\end{subfigure}
\hfill
\begin{subfigure}{0.32\textwidth}
    \includegraphics{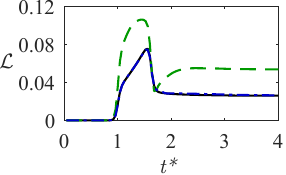}
    \caption{$k = 0.4$, $\Lambda = 30^\circ$}
    \label{fig:LESP_fifth}
\end{subfigure}
\hfill
\begin{subfigure}{0.32\textwidth}
    \includegraphics{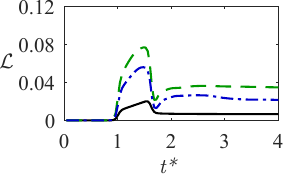}
    \caption{$k = 0.4$, $\Lambda = 60^\circ$}
    \label{fig:LESP_sixth}
\end{subfigure}
\hfill
\begin{subfigure}{1\textwidth}
    \includegraphics{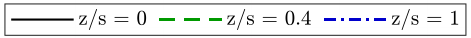}
    \label{fig:LESP_seventh}
\end{subfigure}
  
\caption{Time histories of leading-edge suction parameter magnitude for all cases along the root, 40\% span and tip planes.}
\label{fig:LESP}
\end{figure}

\begin{figure}
\centering

\begin{subfigure}{0.32\textwidth}
    \includegraphics[width=\textwidth]{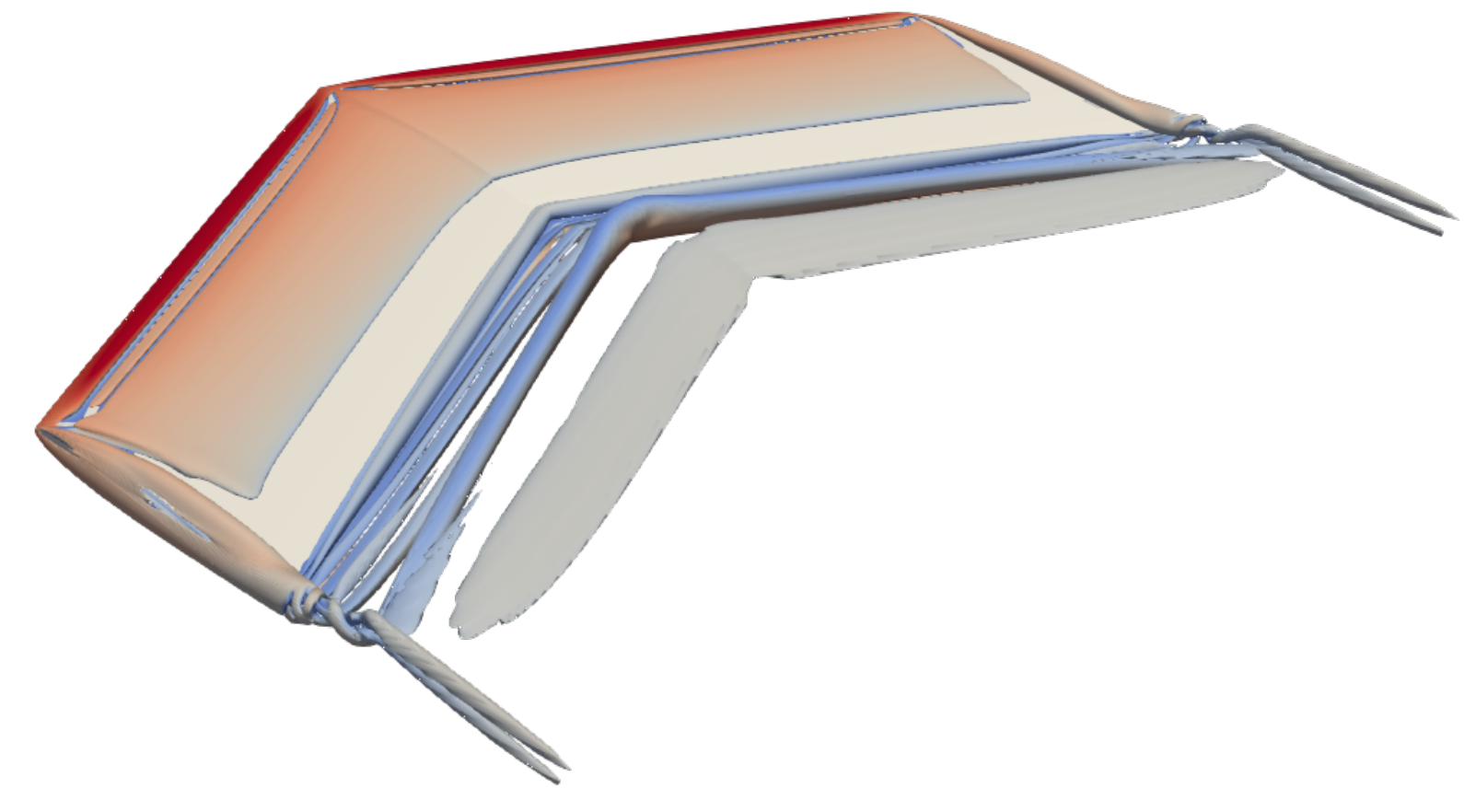}
    \caption{$t^* = 1.4$}
    \label{fig:LESP_Q_first}
\end{subfigure}
\hfill
\begin{subfigure}{0.32\textwidth}
    \includegraphics[width=\textwidth]{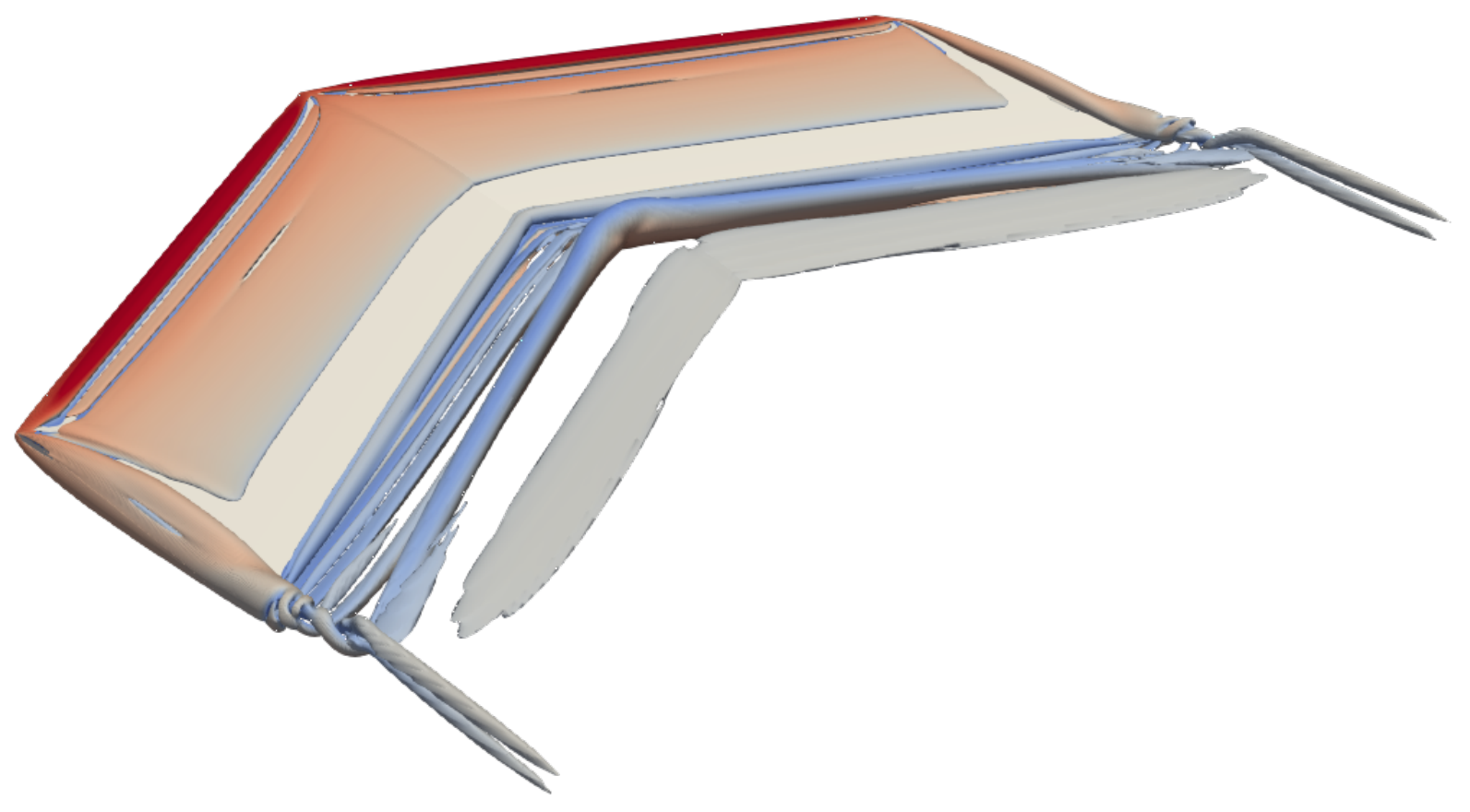}
    \caption{$t^* = 1.44$}
    \label{fig:LESP_Q_second}
\end{subfigure}
\hfill
\begin{subfigure}{0.32\textwidth}
    \includegraphics[width=\textwidth]{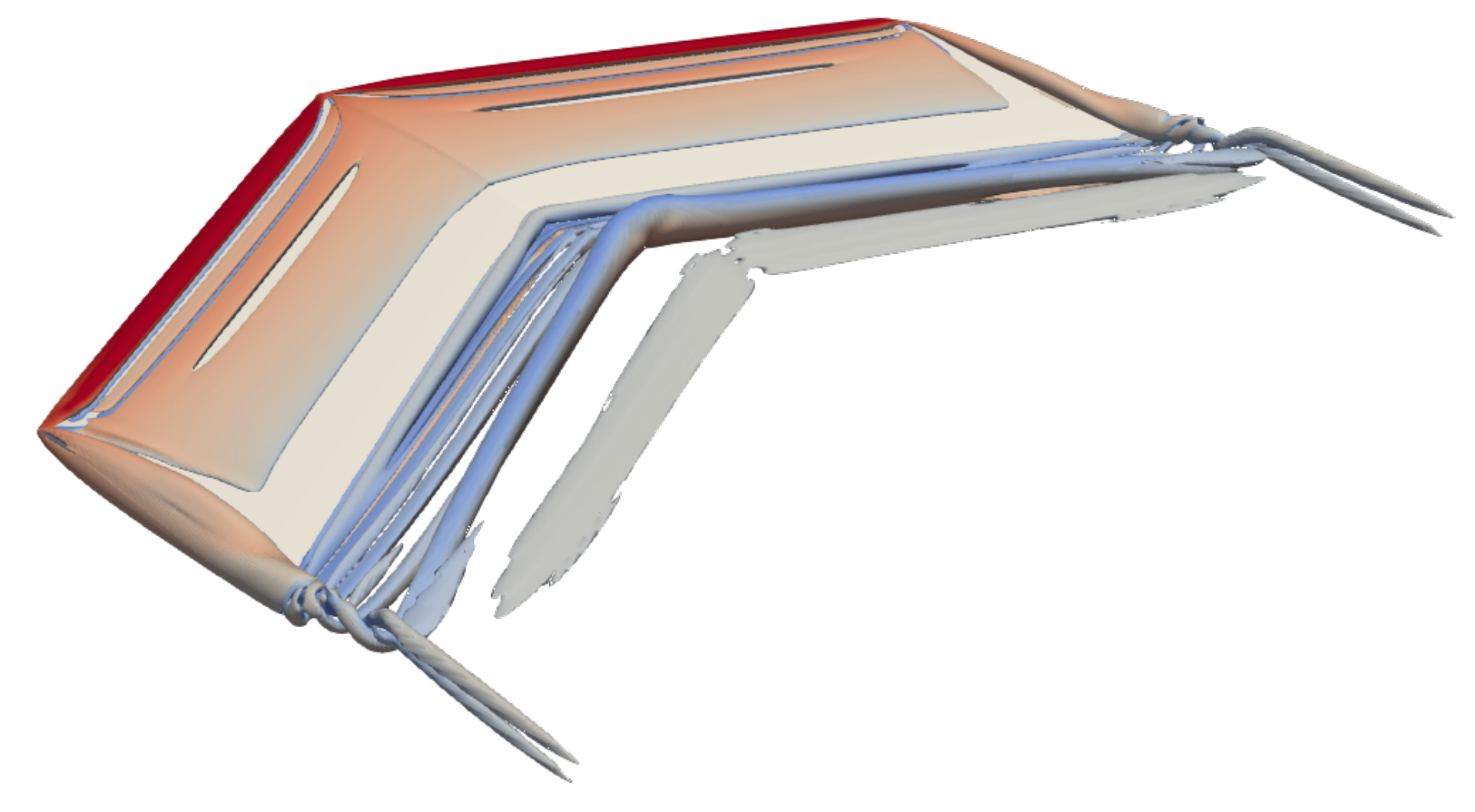}
    \caption{$t^* = 1.48$}
    \label{fig:LESP_Q_third}
\end{subfigure}

\hfill
\begin{subfigure}{1\textwidth}
    \includegraphics{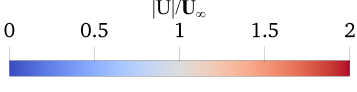}
    \label{fig:vel_mag_scale}
\end{subfigure}

\caption{Isosurfaces of $Q = 1$ colored by normalized velocity magnitude for the $30^\circ$ swept wing at reduced frequency $k = 0.4$ at convective times around 40\% span-predicted LEV initiation (Fig. \ref{fig:LESP_fifth}). The legend also applies to Figs. \ref{fig:Q_1_k_0p05} and \ref{fig:Q_1_k_0p4}.}
\label{fig:LESP_Q}
\end{figure}

To determine LEV initiation, we use the leading-edge suction parameter (LESP) as discussed in section II.A. Fig. \ref{fig:LESP} shows LESP values along three spanwise planes for each wing-motion combination. The root and tip planes are chosen, alongside an arbitrary mid-span plane at 40\% span. To validate the use of the method of Martínez \textit{et al.} \cite{martinez_2022} for swept wing geometries, we compare the quantitative LESP-predicted LEV initiation to qualitative $Q$-criterion isosurfaces in Fig. \ref{fig:LESP_Q}. The moderately swept wing at a high reduced frequency (Fig. \ref{fig:LESP_fifth}) was selected for this analysis as the high reduced frequency results show less spanwise variation in LEV initiation onset, as shown in Fig. \ref{fig:LESP}. The LESP-predicted LEV initiation occurs at $t^* = 1.44$ along the mid-span plane, with the root and tip planes having a delayed maximum LESP value. This results from the outboard spanwise flow at the root and the effect of the tip vortex (TV) delaying peak suction along these planes. The qualitative flow field corroborates this result as this is the convective time at which the shear layer begins to detach from the leading edge near the mid-span and roll up into an LEV. Also, at all times in Fig. \ref{fig:LESP_Q} the shear layer is attached at the root and tip, which agrees with the delayed LEV onset prediction from Fig. \ref{fig:LESP_fifth}. Hence, this analysis demonstrates the validity of the method and generally the use of the LESP concept for swept-wing geometries.

Now considering the general trends displayed in Fig. \ref{fig:LESP}, the most immediate trend is the variation in LESP on each spanwise plane. For example, in Figs. \ref{fig:LESP_first} and \ref{fig:LESP_fourth} the root plane generally has the highest magnitude, whereas for the swept cases it is the mid-span plane. This is a result of no significant spanwise flow outboard from the root being present for the unswept geometry. This is similar between reduced frequencies as the wing geometries govern the spanwise LESP distribution. Also, as the sweep angle increases the peak LESP value decreases, which indicates reducing LEV strength. The root plane LESP magnitude decreases with increasing sweep angle, due to the outboard convection of flow. This pattern is also present to a lesser extent for the mid-span plane.

The low reduced frequency results in Figs. \ref{fig:LESP_first}-\ref{fig:LESP_third} have a general trend of increasing LESP before plateauing near $t^* = 5$, which then reduces after LEV initiation. As the wings plunge, the suction required to keep the shear layer attached to the leading edge increases, until LEV initiation occurs at the peak value. The high reduced frequency results (Figs. \ref{fig:LESP_fourth}-\ref{fig:LESP_sixth}) have a similar general trend, with a much earlier peak due to the motion kinematics. This results in a more abrupt rise and fall in LESP values before and after the peak. There is less spanwise variation in peak LESP location for the same reason as a stronger LEV is induced. This is reflected in the higher peak LESP values displayed across all wing geometries. 

\subsection{Vortex structures}

\begin{figure}
\centering

\begin{subfigure}{0.32\textwidth}
    \includegraphics[width=\textwidth]{Q_1000_unswept_k_0p05.pdf}
    \caption{$k = 0.05$, $\Lambda = 0^\circ$, $t^* = 6$}
    \label{fig:Q_1000_first}
\end{subfigure}
\hfill
\begin{subfigure}{0.32\textwidth}
    \includegraphics[width=\textwidth]{Q_1000_SW30_k_0p05.pdf}
    \caption{$k = 0.05$, $\Lambda = 30^\circ$, $t^* = 6$}
    \label{fig:Q_1000_second}
\end{subfigure}
\hfill
\begin{subfigure}{0.32\textwidth}
    \includegraphics[width=\textwidth]{Q_1000_SW60_k_0p05.pdf}
    \caption{$k = 0.05$, $\Lambda = 60^\circ$, $t^* = 6$}
    \label{fig:Q_1000_third}
\end{subfigure}

\hfill
\begin{subfigure}{0.32\textwidth}
    \includegraphics[width=\textwidth]{Q_1000_unswept_k_0p4.pdf}
    \caption{$k = 0.4$, $\Lambda = 0^\circ$, $t^* = 2$}
    \label{fig:Q_1000_fourth}
\end{subfigure}
\hfill
\begin{subfigure}{0.32\textwidth}
    \includegraphics[width=\textwidth]{Q_1000_SW30_k_0p4.pdf}
    \caption{$k = 0.4$, $\Lambda = 30^\circ$, $t^* = 2$}
    \label{fig:Q_1000_fifth}
\end{subfigure}
\hfill
\begin{subfigure}{0.32\textwidth}
    \includegraphics[width=\textwidth]{Q_1000_SW60_k_0p4.pdf}
    \caption{$k = 0.4$, $\Lambda = 60^\circ$, $t^* = 2$}
    \label{fig:Q_1000_sixth}
\end{subfigure}

\hfill
\begin{subfigure}{1\textwidth}
    \includegraphics{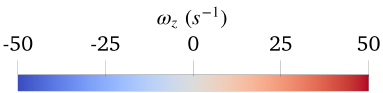}
    \label{fig:Q_1000_seventh}
\end{subfigure}
  
\caption{Isosurfaces of $Q = 1000$ colored by spanwise vorticity ($\omega_z$) for all cases, at convective times where a coherent LEV exists over the wing.}
\label{fig:Q_1000}
\end{figure}

\begin{figure}
\centering

\begin{subfigure}{0.32\textwidth}
    \includegraphics[width=\textwidth]{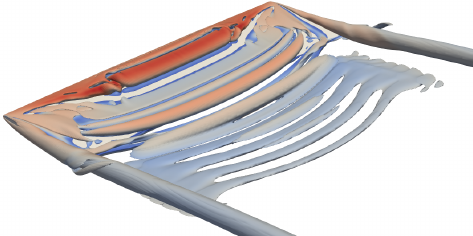}
    \caption{$\Lambda = 0^\circ$, $t^* = 6$}
    \label{fig:Q_1_k_0p05_first}
\end{subfigure}
\hfill
\begin{subfigure}{0.32\textwidth}
    \includegraphics[width=\textwidth]{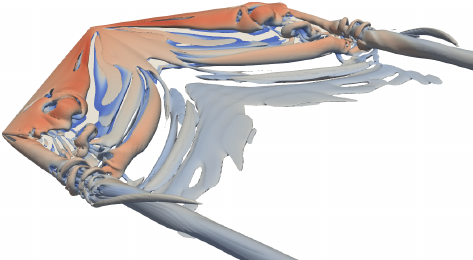}
    \caption{$\Lambda = 30^\circ$, $t^* = 6$}
    \label{fig:Q_1_k_0p05_second}
\end{subfigure}
\hfill
\begin{subfigure}{0.32\textwidth}
    \includegraphics[width=\textwidth]{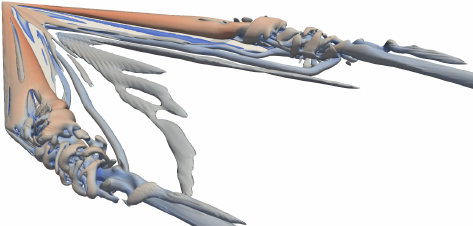}
    \caption{$\Lambda = 60^\circ$, $t^* = 6$}
    \label{fig:Q_1_k_0p05_third}
\end{subfigure}

\hfill
\begin{subfigure}{0.32\textwidth}
    \includegraphics[width=\textwidth]{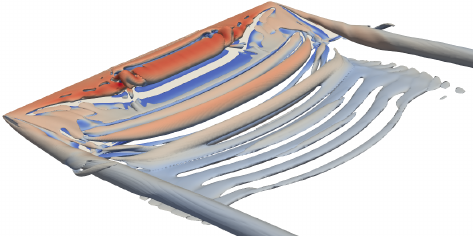}
    \caption{$\Lambda = 0^\circ$, $t^* = 6.2$}
    \label{fig:Q_1_k_0p05_fourth}
\end{subfigure}
\hfill
\begin{subfigure}{0.32\textwidth}
    \includegraphics[width=\textwidth]{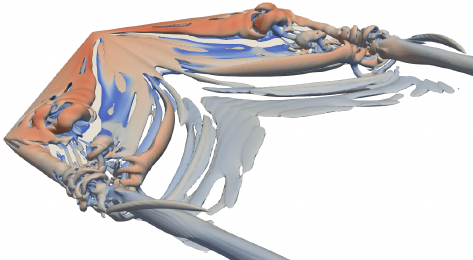}
    \caption{$\Lambda = 30^\circ$, $t^* = 6.2$}
    \label{fig:Q_1_k_0p05_fifth}
\end{subfigure}
\hfill
\begin{subfigure}{0.32\textwidth}
    \includegraphics[width=\textwidth]{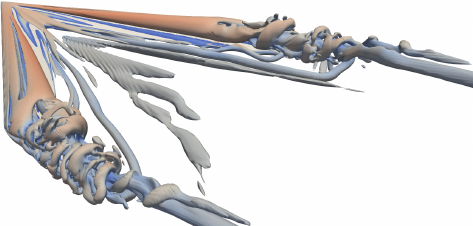}
    \caption{$\Lambda = 60^\circ$, $t^* = 6.2$}
    \label{fig:Q_1_k_0p05_sixth}
\end{subfigure}

\begin{subfigure}{0.32\textwidth}
    \includegraphics[width=\textwidth]{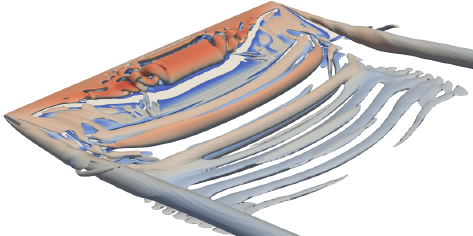}
    \caption{$\Lambda = 0^\circ$, $t^* = 6.4$}
    \label{fig:Q_1_k_0p05_seventh}
\end{subfigure}
\hfill
\begin{subfigure}{0.32\textwidth}
    \includegraphics[width=\textwidth]{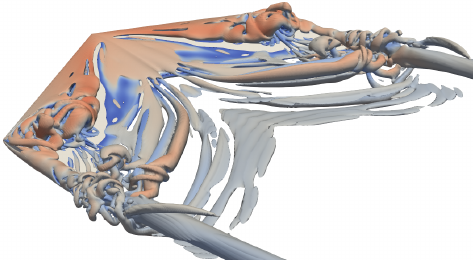}
    \caption{$\Lambda = 30^\circ$, $t^* = 6.4$}
    \label{fig:Q_1_k_0p05_eighth}
\end{subfigure}
\hfill
\begin{subfigure}{0.32\textwidth}
    \includegraphics[width=\textwidth]{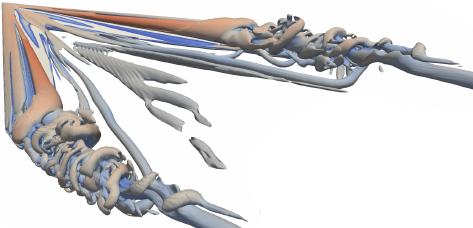}
    \caption{$\Lambda = 60^\circ$, $t^* = 6.4$}
    \label{fig:Q_1_k_0p05_ninth}
\end{subfigure}

\caption{Isosurfaces of $Q = 1$ colored by normalized velocity magnitude for all wing geometries at reduced frequency $k = 0.05$ at convective times where a coherent LEV exists over the wing.}
\label{fig:Q_1_k_0p05}
\end{figure}

\begin{figure}
\centering

\hfill
\begin{subfigure}{0.32\textwidth}
    \includegraphics[width=\textwidth]{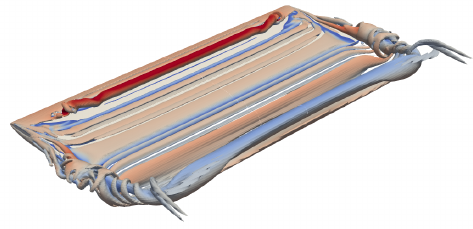}
    \caption{$\Lambda = 0^\circ$, $t^* = 2$}
    \label{fig:Q_1_k_0p4_first}
\end{subfigure}
\hfill
\begin{subfigure}{0.32\textwidth}
    \includegraphics[width=\textwidth]{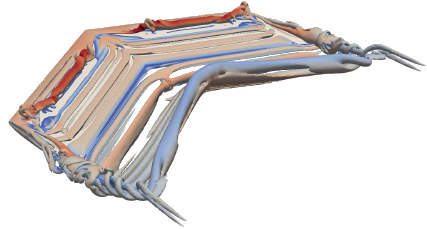}
    \caption{$\Lambda = 30^\circ$, $t^* = 2$}
    \label{fig:Q_1_k_0p4_second}
\end{subfigure}
\hfill
\begin{subfigure}{0.32\textwidth}
    \includegraphics[width=\textwidth]{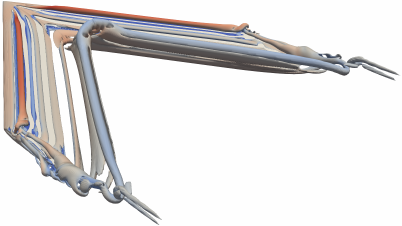}
    \caption{$\Lambda = 60^\circ$, $t^* = 2$}
    \label{fig:Q_1_k_0p4_third}
\end{subfigure}

\begin{subfigure}{0.32\textwidth}
    \includegraphics[width=\textwidth]{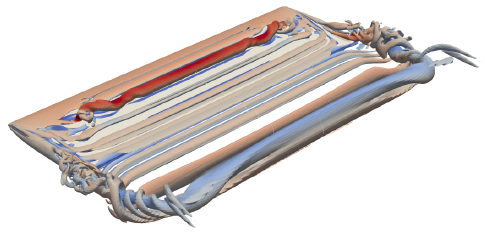}
    \caption{$\Lambda = 0^\circ$, $t^* = 2.2$}
    \label{fig:Q_1_k_0p4_fourth}
\end{subfigure}
\hfill
\begin{subfigure}{0.32\textwidth}
    \includegraphics[width=\textwidth]{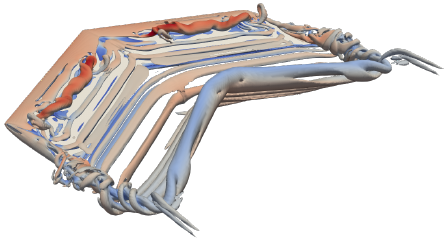}
    \caption{$\Lambda = 30^\circ$, $t^* = 2.2$}
    \label{fig:Q_1_k_0p4_fifth}
\end{subfigure}
\hfill
\begin{subfigure}{0.32\textwidth}
    \includegraphics[width=\textwidth]{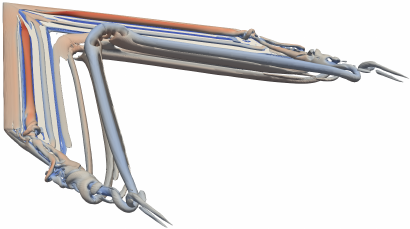}
    \caption{$\Lambda = 60^\circ$, $t^* = 2.2$}
    \label{fig:Q_1_k_0p4_sixth}
\end{subfigure}

\hfill
\begin{subfigure}{0.32\textwidth}
    \includegraphics[width=\textwidth]{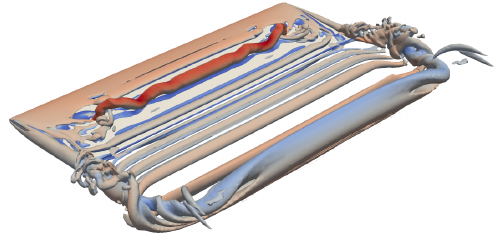}
    \caption{$\Lambda = 0^\circ$, $t^* = 2.4$}
    \label{fig:Q_1_k_0p4_seventh}
\end{subfigure}
\hfill
\begin{subfigure}{0.32\textwidth}
    \includegraphics[width=\textwidth]{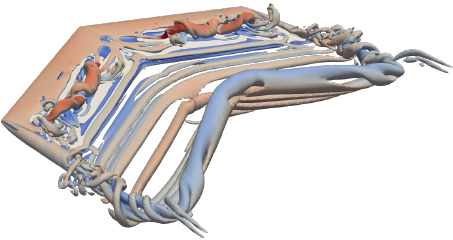}
    \caption{$\Lambda = 30^\circ$, $t^* = 2.4$}
    \label{fig:Q_1_k_0p4_eighth}
\end{subfigure}
\hfill
\begin{subfigure}{0.32\textwidth}
    \includegraphics[width=\textwidth]{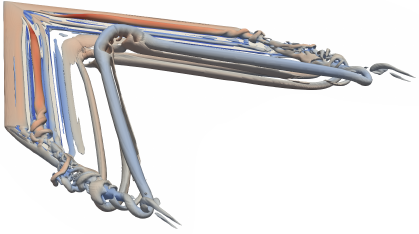}
    \caption{$\Lambda = 60^\circ$, $t^* = 2.4$}
    \label{fig:Q_1_k_0p4_ninth}
\end{subfigure}

\caption{Isosurfaces of $Q = 1$ colored by normalized velocity magnitude for all wing geometries at reduced frequency $k = 0.4$ at convective times where a coherent LEV exists over the wing.}
\label{fig:Q_1_k_0p4}
\end{figure}

We now consider how the vortex structures vary between wing geometries and kinematic conditions, particularly the LEV and tip vortex (TV) cores in Fig. \ref{fig:Q_1000}, and more generally in Figs. \ref{fig:Q_1_k_0p05} and \ref{fig:Q_1_k_0p4}. This is important as the formation of vortices in the vicinity of the wing can strongly affect the aerodynamic loading of the wing, as will be demonstrated in section IV.E.

Fig \ref{fig:Q_1000} shows that there is greater variation in vortex structures between sweep angles as opposed to reduced frequency. The plunge rate primarily governs the strength of the vortices and whether they remain attached, whereas the geometries determine the shape and characteristic features. The unswept flow fields in Figs. \ref{fig:Q_1000_first} and \ref{fig:Q_1000_fourth} both display a straight LEV parallel to the leading edge over the inboard third of the span. A spiral vortex leg structure is located outboard of this for the low reduced frequency case (Fig. \ref{fig:Q_1000_first}), which is connected to the shear layer towards the tip region. Contrastingly, the LEV remains parallel to the leading edge over a much greater spanwise extent in the high reduced frequency case (Fig. \ref{fig:Q_1000_fourth}) as a stronger vortex is formed due to the faster motion kinematics. A twisting vortex leg region is also apparent, however this terminates over the wing surface. The TV produced is much longer for the high reduced frequency case, due to the stronger vorticity generated by the greater plunge rate.

The moderately swept wings (Figs. \ref{fig:Q_1000_second} and \ref{fig:Q_1000_fifth}) exhibit vastly different vortex structures. First, the low reduced frequency case in Fig. \ref{fig:Q_1000_second} has an attached LEV towards the root, which has undergone breakdown in the outboard regions as too weak vorticity was fed into the LEV core to sustain the coherent LEV structure in the outboard regions, due to the slow wing acceleration. The TV has similar unsteady regions downstream of the mid-chord. However, for the high reduced frequency case (Fig. \ref{fig:Q_1000_fifth}), the LEV is detached from the leading edge and a coherent structure is present across the span. A pair of vortex legs are present inboard and outboard of the central LEV core, alongside an elongated TV, both of which are similar to that in Fig. \ref{fig:Q_1000_fourth}. The high reduced frequency LEV is more stable as a stronger LEV is induced by the faster kinematics, as shown in Fig. \ref{fig:LESP}.

Finally, the highly swept wings in Figs. \ref{fig:Q_1000_third} and \ref{fig:Q_1000_sixth} both have a straight LEV, however in the low reduced frequency case (Fig. \ref{fig:Q_1000_third}) this is at an angle to the leading-edge due to remaining attached near the root. The LEV also terminates near the mid-span for this case as it is weak and breaks down. There are some unsteady features near the tip trailing edge, however no coherent TV exists. For the high reduced frequency case (Fig. \ref{fig:Q_1000_sixth}) the stronger, straight LEV is parallel to the leading edge and present over the majority of the span. Towards the root, the LEV bends towards the leading edge where the shear layer has detached. Near the tip, the LEV interacts with and displaces the TV.

Isosurfaces of $Q = 1$ are now used to visualize the detailed flow field captured by the IDDES methodology and analyse the LEV and TV outside of the core flow regions. By defining the $Q$ value to be such that regions where vorticity magnitude equals the strain rate magnitude, additional regions of lower vorticity, such as the trailing-edge vortices (TEVs), can be analyzed. The flow fields in Figs. \ref{fig:Q_1_k_0p05} and \ref{fig:Q_1_k_0p4} show the LEV after formation and the subsequent convection away from the leading-edge for the low and high reduced frequency cases respectively. 

First considering the unswept wing at low reduced frequency in Figs. \ref{fig:Q_1_k_0p05_first}, \ref{fig:Q_1_k_0p05_fourth} and \ref{fig:Q_1_k_0p05_seventh} the most apparent feature is the instability of the LEV propagating inboard as time progresses. This feature is due to the instability of the attached leading-edge shear layer near the wing tips. The central LEV core also thickens as it is pushed towards the root, observing the conservation of angular momentum. The TV remains steady throughout the motion and weak TEVs are formed. There is a strong region of vorticity just past the trailing edge that resulted from a shear layer that did not fully roll up into an LEV.

The main difference between the unswept and swept wing geometries is the formation of a co-rotating vortex pair about the root due to the sharp apex of the wing. The moderately swept wing in Figs. \ref{fig:Q_1_k_0p05_second}, \ref{fig:Q_1_k_0p05_fifth} and \ref{fig:Q_1_k_0p05_eighth} exhibits LEV breakdown that again propagates inboard. However in this case, breakdown initiates from LEV bursting in the outboard regions, as shown in Fig. \ref{fig:Q_1_k_0p05_second}. A small arch vortex is formed where this instability originates, which grows as the breakdown progresses. The LEV is also still connected to the leading-edge shear layer in inboard regions due to the low rate of motion. TV breakdown initiates near the trailing edge due to the instability caused by the LEV, which progresses downstream with time. Fewer weak TEVs are formed compared to the unswept case and a similar region of strong vorticity from the previous shear layer that is displaced upstream towards the root is present. Both of these effects are due to the shape of the trailing-edge geometry.

The highly swept wing in Figs. \ref{fig:Q_1_k_0p05_third}, \ref{fig:Q_1_k_0p05_sixth} and \ref{fig:Q_1_k_0p05_ninth} has a stable LEV over most of the span for all times. This is due to the high sweep angle turning the incident flow about a smaller angle than the other geometries, hence the flow can more readily remain attached. The LEV remains connected at the root and also near the tip. However, no apparent TV and large-scale unsteady features are observed in the outboard regions. This grows with time, but cannot propagate appreciably inboard due to the high sweep angle as it has to travel primarily upstream. Weak TEV structures are also present in the wake. 

Now considering the high reduced frequency cases in Fig. \ref{fig:Q_1_k_0p4}, the flow field is significantly different to the low reduced frequency case for all wing geometries. The coherent LEV is detached from the leading edge and convects downstream and much stronger TEVs are formed, due to the impulse-like wing acceleration. The TEVs are more noticeably deflected upwards for the swept wing geometries, although this is also present for the unswept wing. The strongest TEV located furthest downstream is formed during the initial wing acceleration. Subsequent weaker TEVs are formed during the plunge motion.

The unswept wing in Figs. \ref{fig:Q_1_k_0p4_first}, \ref{fig:Q_1_k_0p4_fourth} and \ref{fig:Q_1_k_0p4_seventh} features a narrow LEV, with spiral-type instabilities from the LEV legs propagating inboard as time progresses. The legs themselves also move inboard as vorticity is shed from the tip inboard, unlike for the swept cases where vorticity is shed outboard from the root. Similarly to the low reduced frequency case, the LEV also thickens near the root. The TV re-attaches to the wing tip as time progresses, which was initially disrupted by the impulse-like deceleration of the wing to a constant plunge rate at $t^* = 1.6$. Downstream of the trailing edge, the TV is chaotic.

Over the moderately swept wing (Figs. \ref{fig:Q_1_k_0p4_second}, \ref{fig:Q_1_k_0p4_fifth} and \ref{fig:Q_1_k_0p4_eighth}) the LEV vortex pair has strong instabilities originating from the root and tip LEV legs, that propagate towards the mid-span. Therefore, the LEV becomes a spiral structure along its length as time progresses. Compared to the low reduced frequency case (Figs. \ref{fig:Q_1_k_0p05_second}, \ref{fig:Q_1_k_0p05_fifth} and \ref{fig:Q_1_k_0p05_eighth}), LEV breakdown initiates due to leg instabilities, rather than LEV bursting due to the differences in kinematics. The detached TV reattaches, similarly to the unswept wing and the downstream wake is highly unsteady.

Finally, the highly swept wing in Figs. \ref{fig:Q_1_k_0p4_third}, \ref{fig:Q_1_k_0p4_sixth} and \ref{fig:Q_1_k_0p4_ninth} has a straight LEV over most of the spanwise extent at all times due to the shape of the geometry. The LEV remains attached near the root, hence additional vorticity is fed into the LEV, strengthening it. This can be observed by the thicker LEV regions towards the root in Figs. \ref{fig:Q_1_k_0p4_sixth} and \ref{fig:Q_1_k_0p4_ninth}. Towards the tip, the LEV interacts with the TV and displaces it from the leading edge. Hence, there are no LEV legs over this wing planform. Instability due to the interaction of the two vortex structures increases with time, although this is mainly confined to the TV wake. 

\subsection{LEV convection rate}

\begin{figure}
\centering

\begin{subfigure}{1\textwidth}
    \includegraphics{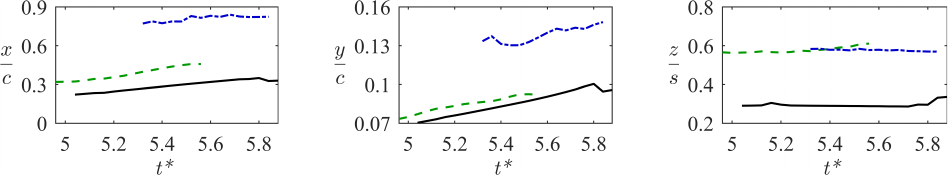}
    \caption{$k = 0.05$}
    \label{fig:LEV_3d_first}
\end{subfigure}
\hfill
\begin{subfigure}{1\textwidth}
    \includegraphics{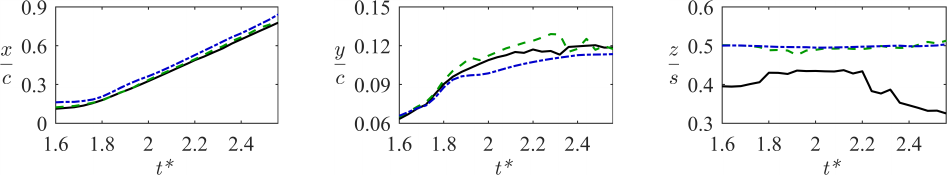}
    \caption{$k = 0.4$}
    \label{fig:LEV_3d_second}
\end{subfigure}
\caption{Normalized chordwise (left), wall-normal (centre) and spanwise (right) locations of the LEV centroid for convective times after LEV initiation until breakdown.}
\label{fig:LEV_3d}
\end{figure}

To understand the motion of the LEV after initiation, the LEV centroid location is decomposed into its Cartesian coordinate components to assess the convection rate and is shown in Fig. \ref{fig:LEV_3d}. For the low reduced frequency cases in Fig. \ref{fig:LEV_3d_first}, the formation time of the coherent LEV differs between geometries (as shown in Figs. \ref{fig:LESP_first}-\ref{fig:LESP_third}). Hence, we plot the times at which a coherent LEV exists over the wing for each case. 

The chord-wise LEV location varies significantly between geometries for the low reduced frequency case (Fig. \ref{fig:LEV_3d_first}), unlike for the high reduced frequency case. This is due to the attached nature of the LEV over these wings, as demonstrated in Fig. \ref{fig:Q_1_k_0p05}. As the sweep angle increases, the LEV is located further downstream, with a larger difference between the two swept wing geometries. This is due to the attached nature of these LEVs forming an angle relative to the wing sweep angle, as shown in Figs. \ref{fig:Q_1000_second} and \ref{fig:Q_1000_third}. The LEV convects at a near-linear rate in the chord-wise direction for all wing-kinematics combinations, however with a region of near-stationary LEV location for the high reduced frequency cases until $t^* = 1.8$. This results from the LEV remaining attached to the leading edge for a short time, before convecting downstream. The LEV convects at a much faster rate at a high reduced frequency after this point as the LEV is detached from the leading edge. 

For the low reduced frequency cases, the wall-normal location shows a similar overall trend to the chord-wise location; as the LEV is located further downstream it is also displaced higher away from the chord-line. For each geometry, the increase in wall-normal location increases at a similar rate that is mostly linear, except for the highly swept wing as the LEV is located much further above the chord-line. At high reduced frequency, the wall-normal location shows significant variation between geometries when compared to the chord-wise location, albeit while being an order of magnitude smaller. The moderately swept wing generally has the greatest separation from the chord-line, due to the spiral nature of the LEV and the highly swept wing the least, as it is the most straight and aligned with the leading edge. The same general trend is observed for all three wing geometries that appear to asymptotically approach $\frac{y}{c} = 0.12$ as the LEV diffuses away to this extent.

The spanwise LEV location is considerably different between the unswept and swept wing geometries due to the vortex pair generated about the wing apex in the swept wing cases. Hence, for the high reduced frequency cases, the LEV is centered about the mid-span for the swept wing cases and slightly outboard of this for the low reduced frequency cases. This is due to the attached LEV towards the root making the LEV thicker in the outboard regions as vorticity is fed into the LEV for the low reduced frequency cases. The unswept low reduced frequency LEV is located much further inboard due to the thin attached shear layer from the tip outboard of the LEV, as shown in Figs. \ref{fig:Q_1000_first} and \ref{fig:Q_1_k_0p05_first}. For the high reduced frequency case, the unswept wing LEV tends to move inboard after $t^* = 2.2$, due to the inboard movement of the LEV legs, as shown in Figs. \ref{fig:Q_1_k_0p4_fourth} and \ref{fig:Q_1_k_0p4_seventh}.

\subsection{LEV lift distribution}

\begin{figure}
\centering

\begin{subfigure}{0.32\textwidth}
    \includegraphics{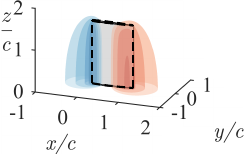}
    \caption{$\Lambda = 0^\circ$}
    \label{fig:phi_first}
\end{subfigure}
\hfill
\begin{subfigure}{0.32\textwidth}
    \includegraphics{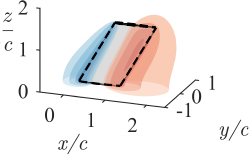}
    \caption{$\Lambda = 30^\circ$}
    \label{fig:phi_second}
\end{subfigure}
\hfill
\begin{subfigure}{0.32\textwidth}
    \includegraphics{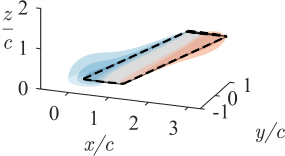}
    \caption{$\Lambda = 60^\circ$}
    \label{fig:phi_third}
\end{subfigure}

\hfill
\begin{subfigure}{1\textwidth}
    \includegraphics{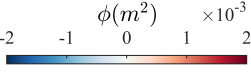}
    \label{fig:phi_legend}
\end{subfigure}

\caption{Lift influence potentials ($\phi$) for each wing geometry, with the wing boundaries indicated by the dashed lines.}
\label{fig:phi}
\end{figure}

As discussed in section II.B, we construct an influence potential ($\phi$) to enable quantitative partitioning of contributions to lift. $\phi$ describes the spatial influence of the wing geometries on the resultant lift due to vorticity. Fig. \ref{fig:phi} shows the $\phi$ fields for the three geometries considered in this study. For all sweep angles, $\phi$ is symmetric about the chord-line ($\frac{y}{c} = 0$) as it is a purely geometrical quantity. 

For the unswept wing, $\phi$ is zero along $\frac{x}{c} \approx 0.3$. The magnitude of $\phi$ is greatest at the wing leading edge and reduces away from the surface. The positive peak is shifted upstream of the trailing edge and shows asymmetry compared to the leading edge distribution. This is due to the difference in geometry between the rounded leading edge and sharp trailing edge of the wing \cite{zhu&breuer_2023}. In a spanwise sense, the size of $\phi$ contours reduces towards the wing tip due to the sharp geometrical feature.

For the $30^\circ$ swept case, the positive $\phi$ region has a considerably larger volume than the unswept case and the negative $\phi$ region has a slightly smaller volume due to the swept geometry. The plane about which $\phi = 0$ persists, however near the tip this is skewed towards the leading edge. The same overall trend is present, with peak $\phi$ persisting further towards the tip near the trailing edge and less so at the leading edge compared to the unswept geometry. The positive $\phi$ contours also thicken towards the mid-span before reducing towards the tip.

The most highly swept case has the $\phi = 0$ plane further shifted towards the leading edge near the tip, and there is also more significant deformation away from the leading edge near the root as the sweep angle is more extreme. The volume of the positive $\phi$ region is much smaller than either of the other two geometries and the peak $\phi$ magnitude is also smaller. The shape of the contours bulges towards the tip and root for the positive and negative $\phi$ regions respectively.

\begin{figure}
\centering

\begin{subfigure}{0.32\textwidth}
    \includegraphics[width=\textwidth]{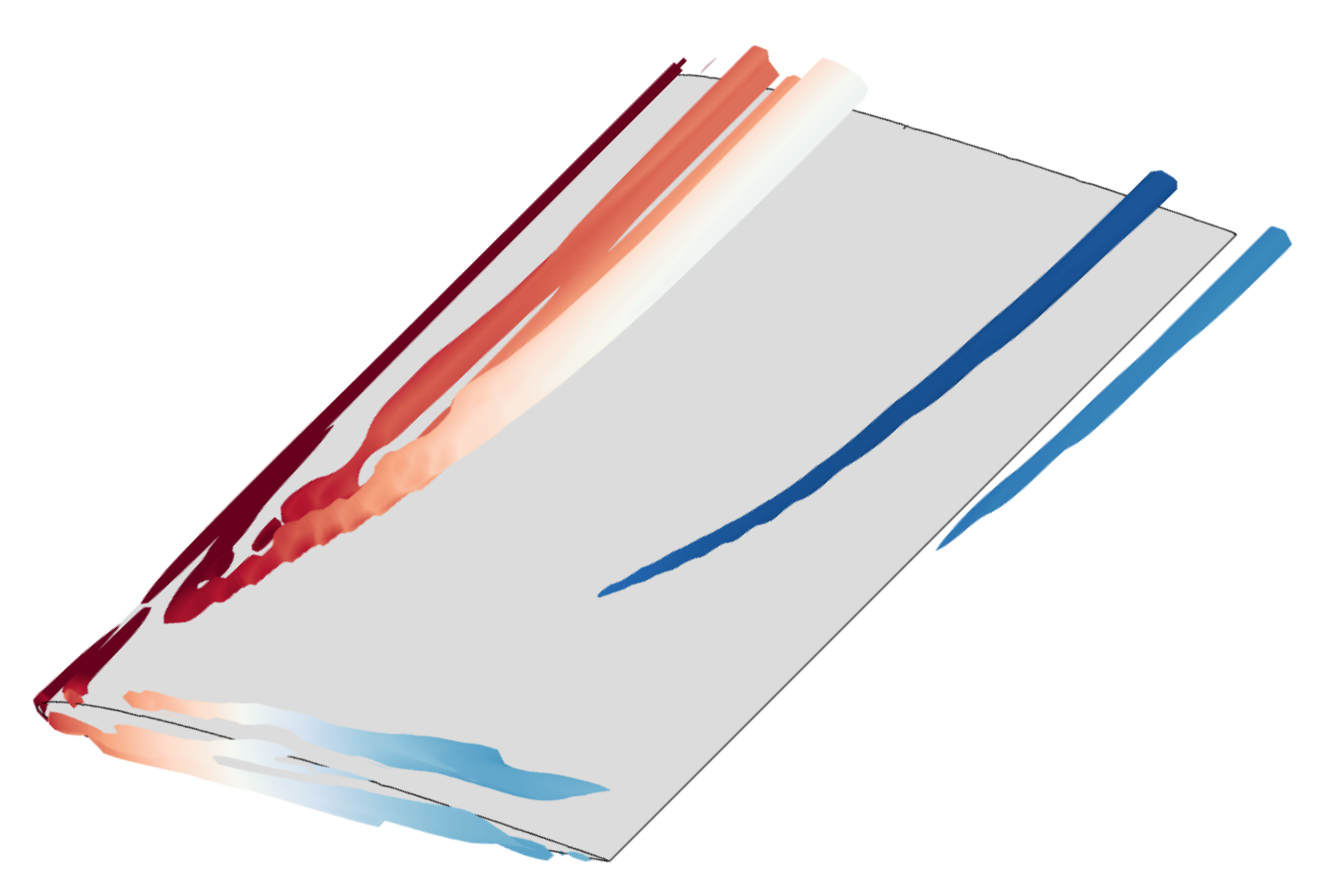}
    \caption{$k = 0.05$, $\Lambda = 0^\circ$, $t^* = 5.6$}
    \label{fig:lift_density_dist_first}
\end{subfigure}
\hfill
\begin{subfigure}{0.32\textwidth}
    \includegraphics[width=\textwidth]{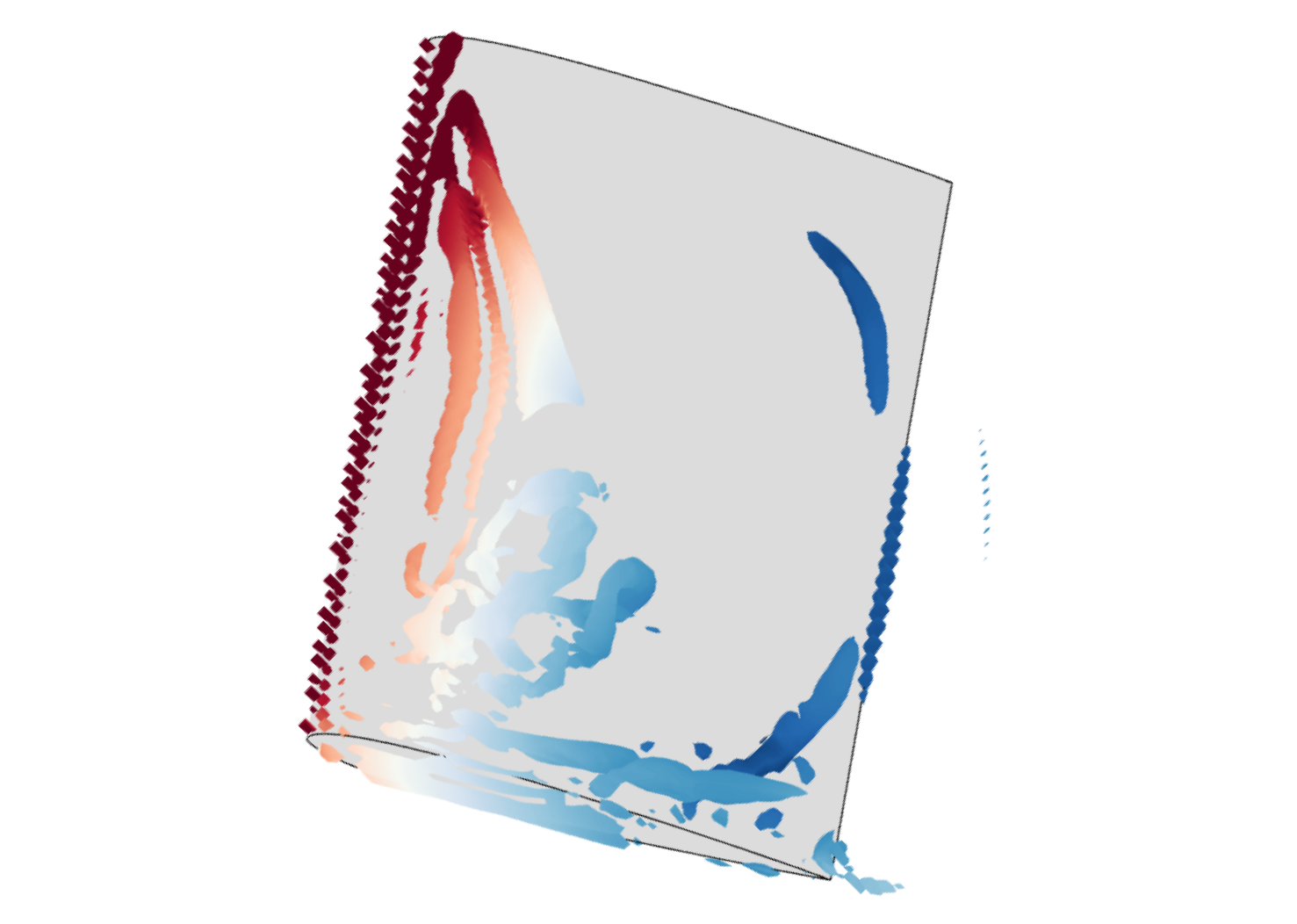}
    \caption{$k = 0.05$, $\Lambda = 30^\circ$, $t^* = 5.6$}
    \label{fig:lift_density_dist_second}
\end{subfigure}
\hfill
\begin{subfigure}{0.32\textwidth}
    \includegraphics[width=\textwidth]{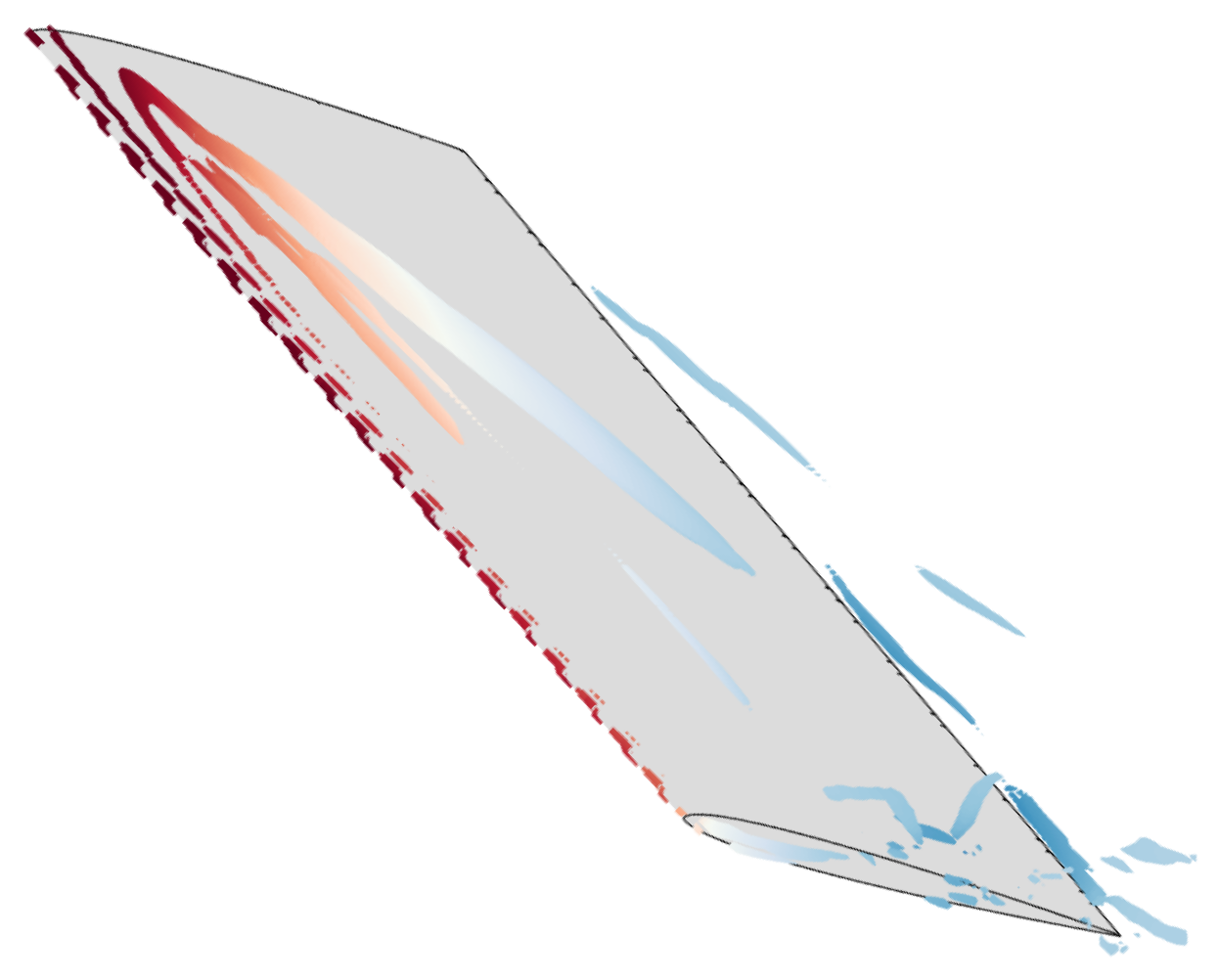}
    \caption{$k = 0.05$, $\Lambda = 60^\circ$, $t^* = 5.6$}
    \label{fig:lift_density_dist_third}
\end{subfigure}

\hfill
\begin{subfigure}{0.32\textwidth}
    \includegraphics[width=\textwidth]{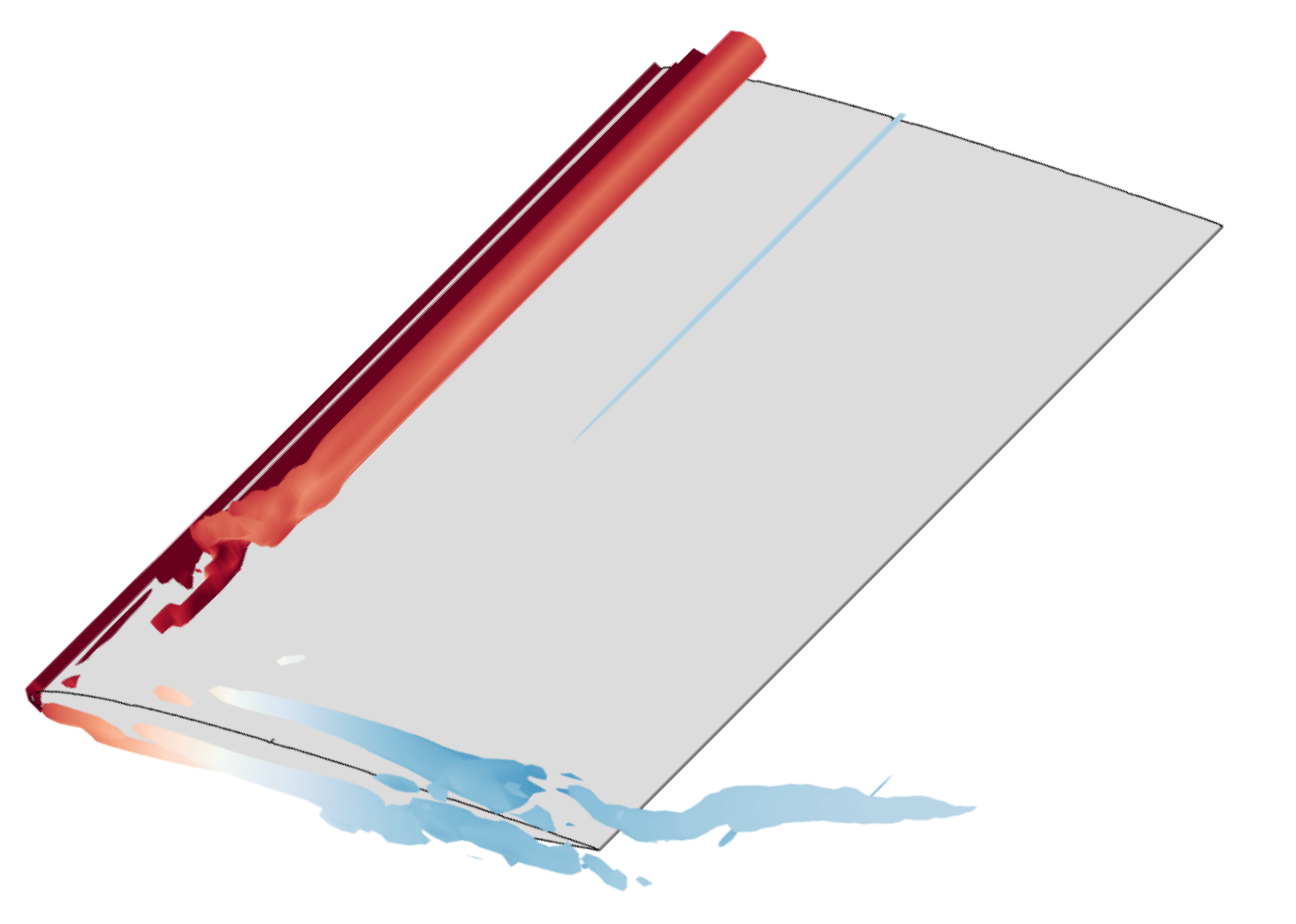}
    \caption{$k = 0.4$, $\Lambda = 0^\circ$, $t^* = 1.8$}
    \label{fig:lift_density_dist_fourth}
\end{subfigure}
\hfill
\begin{subfigure}{0.32\textwidth}
    \includegraphics[width=\textwidth]{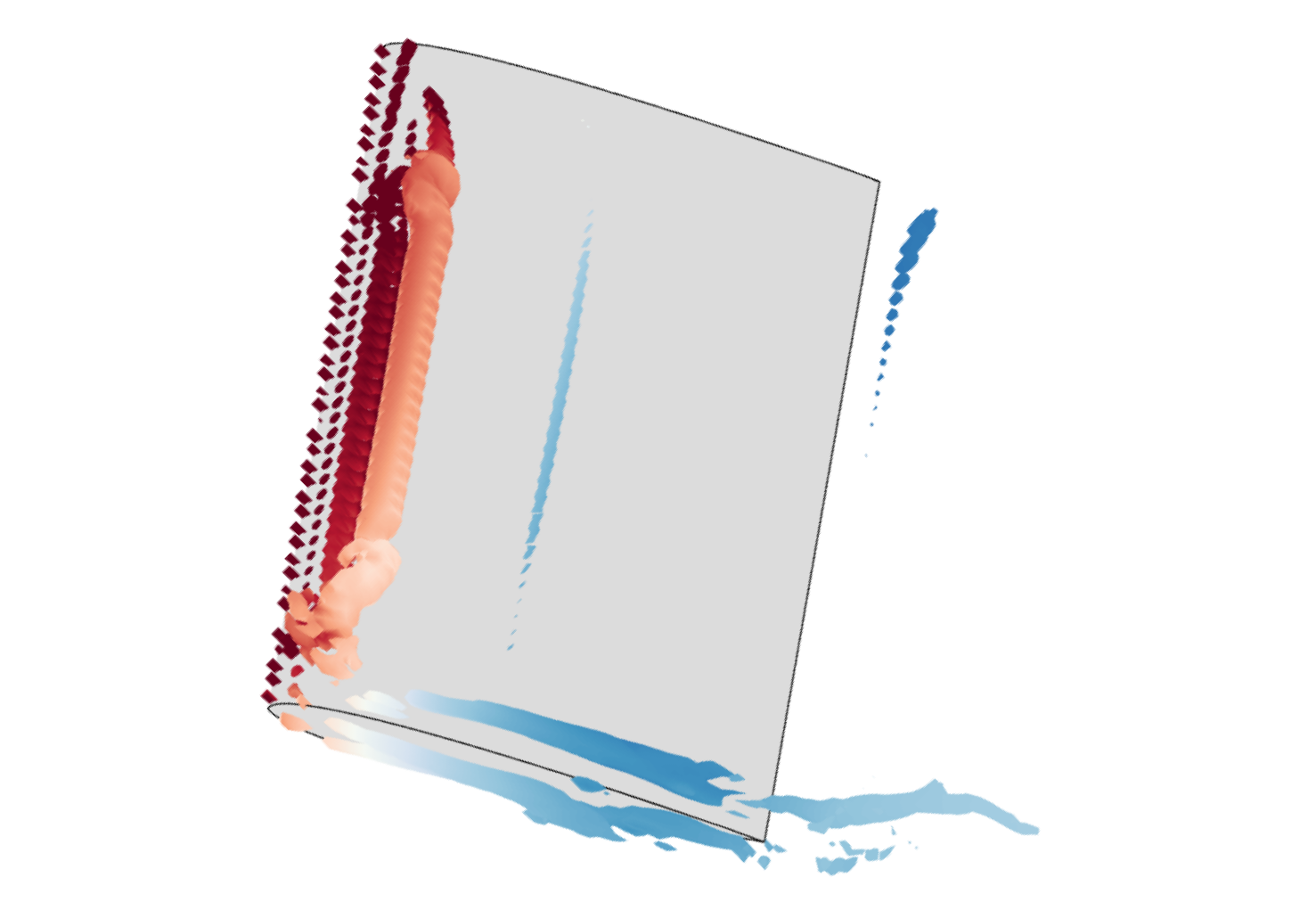}
    \caption{$k = 0.4$, $\Lambda = 30^\circ$, $t^* = 1.8$}
    \label{fig:lift_density_dist_fifth}
\end{subfigure}
\hfill
\begin{subfigure}{0.32\textwidth}
    \includegraphics[width=\textwidth]{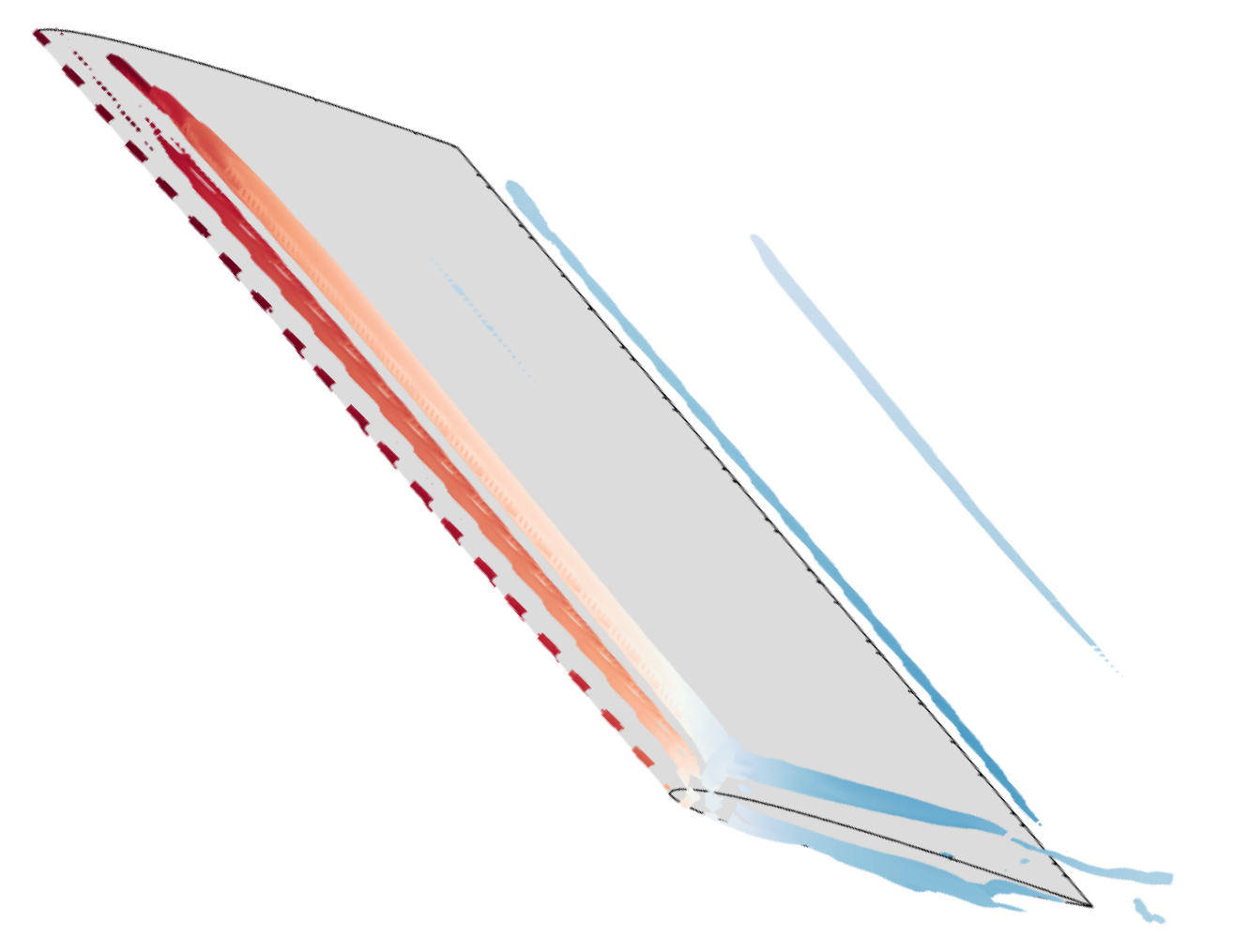}
    \caption{$k = 0.4$, $\Lambda = 60^\circ$, $t^* = 1.8$}
    \label{fig:lift_density_dist_sixth}
\end{subfigure}

\hfill
\begin{subfigure}{1\textwidth}
    \includegraphics{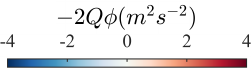}
    \label{fig:lift_density_dist_legend}
\end{subfigure}
  
\caption{Isosurfaces of $Q = 1000$ colored by lift density distribution ($-2Q\phi$) for all cases, at convective times where a coherent LEV exists over the wing.}
\label{fig:lift_density_dist}
\end{figure}

\begin{figure}
\centering
\begin{subfigure}{0.49\textwidth}
    \includegraphics{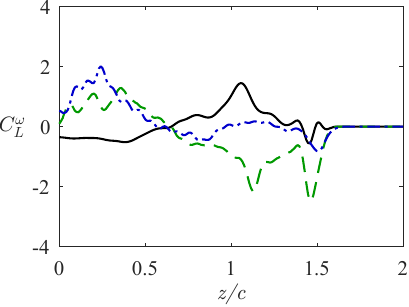}
    \caption{$k = 0.05$}
    \label{fig:C_L_omega_first}
\end{subfigure}
\hfill
\begin{subfigure}{0.49\textwidth}
    \includegraphics{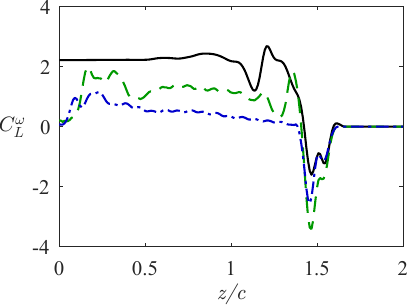}
    \caption{$k = 0.4$}
    \label{fig:C_L_omega_second}
\end{subfigure}
\caption{Spanwise vorticity-induced lift distribution for all cases.}
\label{fig:C_L_omega}
\end{figure}

To obtain the lift density distribution, $-2Q\phi$, we multiply the $\phi$ field by the $Q$-field extracted from simulation data. The resolution of the interpolated $Q$-field is $0.02c$ in the chord-wise and wall-normal directions and $0.01c$ in the spanwise direction. Numerical artifacts due to the resolution of the interpolation are present across the span, particularly for the moderately swept wing. We visualize the contribution to lift of the LEV and TV by coloring $Q = 1000$ isosurfaces by $-2Q\phi$ in Fig. \ref{fig:lift_density_dist}. The effects on lift contribution discussed below are due to the shapes of the influence fields, hence the geometry, as we plot a single value of $Q$. We additionally plot the spanwise distribution of this lift in Fig. \ref{fig:C_L_omega}, which is calculated by integrating over spanwise planes as in equation \ref{vorticity_force}.

At low reduced frequency, the LEV over the unswept wing (Fig. \ref{fig:lift_density_dist_first}) has a weak positive contribution to lift, which becomes stronger as the thin shear layer travels towards the leading edge in the outboard regions. The TV has a positive contribution near the leading edge that transitions to become negative further downstream. This is due to the negative $\phi$ region for the unswept wing over the tip. The moderately swept wing (Fig. \ref{fig:lift_density_dist_second}) has an LEV that strongly contributes to lift near the root, reducing along its spanwise extent until it becomes negative near the mid-span. Outboard of this, the unsteady downstream features primarily have a negative lift contribution. The TV also contributes negatively to lift. The most highly swept wing (Fig. \ref{fig:lift_density_dist_third}) has an LEV that again contributes positively to lift near the root, which becomes negative in the outboard regions. The negative region is of a greater magnitude and much larger than the moderately swept wing. The unsteady wake structures near the tip trailing edge have a negative lift contribution. 

At high reduced frequency, the unswept geometry (Fig. \ref{fig:lift_density_dist_fourth}) has a straight LEV, parallel to the leading edge that has a strong positive contribution to lift. The TV contributes similarly to the low reduced frequency case; positive near the leading edge, then negative downstream. The moderately swept wing (Fig. \ref{fig:lift_density_dist_fifth}) has an LEV that also contributes positively to lift, albeit less strongly than the unswept wing. The magnitude of this lift also reduces towards the tip, due to the shape of the influence field. This analysis similarly applies to the highly swept wing in Fig. \ref{fig:lift_density_dist_sixth}. The TVs for both swept-wing geometries primarily contribute negatively to lift, as traditional theory predicts.

The spanwise lift distribution for the low reduced frequency cases (Fig. \ref{fig:C_L_omega_first}) shows that near $\frac{z}{c} = 1$, the unswept wing has a strong positive lift spike. This is caused by the attached shear layer bending upstream into a region of more negative $\phi$. This is also just outboard of the region of negative vorticity near the trailing edge of the wing, which causes the lift contribution to be negative overall over the inboard 50\% of the span despite the positive LEV contribution. There is only a small negative spike due to the TV at the tip. By contrast, the moderately swept wing has a series of positive lift peaks near the root, due to the LEV, which drops to be negative outboard of $\frac{z}{c} = 0.7$. This coincides approximately with the breakdown location of the LEV. A broad region of negative lift then exists over the outboard regions, with two prominent spikes. The first results from a hairpin vortex over the wing and the second from the TV. Finally, the highly swept wing has a stronger and broader region of positive lift near the root than the moderately swept wing, which becomes negative at the same spanwise location. However, due to the lack of transient LEV features across the outboard regions of the span, the magnitude of the negative lift is much smaller. The negative spike at the wing tip is small due to minimal vortex structures.

Finally, the spanwise lift distribution at the high reduced frequency shows a strong positive lift contribution for the unswept wing over most of the span; see Fig. \ref{fig:C_L_omega_second}. Up to $\frac{z}{c} = 1$, the lift is almost constant, due to the straight LEV aligned to the leading edge. Where the LEV leg is present, there is a lift fluctuation, before a small negative lift spike due to the TV. As the sweep angle increases, the contribution to lift decreases. The moderately swept wing has a broad positive lift peak towards the root. This is due to the region of strong vorticity at the base of the leg. Outboard of this there is a drop in lift due to instability caused by the LEV leg. Then the lift steadily reduces, while still being positive due to the LEV. Near the outboard LEV leg location, the lift fluctuates similarly to the unswept geometry. The TV-induced negative lift is the greatest magnitude of all the geometries due to the lack of positive lift contribution for the unswept wing, combined with the stronger TV generated than the highly swept wing. The most highly swept wing has a similar lift peak towards the root to the moderately swept case, however it is smaller in magnitude due to the weaker shear layer vorticity upstream of the LEV. Lift then steadily reduces towards the tip, as shown in Fig. \ref{fig:lift_density_dist_sixth} becoming near-zero before the TV-induced negative lift. The negative lift peak is approximately halfway in magnitude between the spikes for the other geometries.

\section{Conclusions}

We have investigated the effects of sweep angle and reduced frequency on the three-dimensional nature of the LEV over plunging wings using high-fidelity IDDES. By focusing on the sweep angle, this study aimed to further understand how biological fliers attain agile flight dynamics by manipulating the LEV. The effects of reduced frequency were also studied to investigate unsteady aerodynamic effects.

The lift coefficient has a non-linear decrease as the sweep angle increases during the ramp-down motion for both the reduced frequencies. At the high reduced frequency, large added-mass spikes in lift were present due to the impulse-like acceleration of the wing.

The use of LESP and the method of Martínez \textit{et al.} \cite{martinez_2022} were validated by comparing the quantitative LESP-predicted LEV initiation onset to qualitative flow field visualization via $Q$-criterion isosurfaces. LESP values show that a stronger LEV is initiated at a higher reduced frequency. Additionally, unswept wings have the highest LESP magnitude along the root plane. Whereas over swept wings peak LESP magnitude occurs along the mid-span plane due to the sharp wing geometry at the apex of the swept wings, hence vorticity is shed outboard.

Regarding the detailed vortex structures surrounding the wings highlighted by the IDDES methodology, the most significant difference is the formation of a co-rotating vortex pair about the wing apex over the swept wing geometries. At high reduced frequency, the stronger LEV detaches from the leading edge, convecting downstream, with  LEV legs forming over the unswept and moderately swept wings. Stronger TEVs and an unsteady TV are present at high reduced frequency due to the faster rate of motion. LEV breakdown initiates from LEV leg instabilities at a high reduced frequency and vortex bursting at a low reduced frequency respectively.

The LEV primarily convects in the chord-wise direction, at a much faster rate at a high reduced frequency once the LEV has detached from the leading edge. The height of the LEV above the wing approaches a constant value at a high reduced frequency for all wing geometries. The LEV center is located much further inboard for the unswept wing as there is a single LEV across the span.

To understand the spatial distribution of LEV-induced lift, we used the FPM \cite{menon&mittal_2021,menon&mittal_2021b} to isolate lift due to vorticity. The sweep angle affects how vorticity influences lift as the influence field ($\phi$) depends only on the wing geometry and location. The stronger LEV induces more lift at a high reduced frequency. LEV-induced lift decreases as the sweep angle increases and the  LEV contributes more negatively to lift towards the wing tip at a high reduced frequency. The unswept wings have a constant lift contribution over inboard regions of the span at both reduced frequencies. Whereas, both swept wings have an LEV-induced lift peak near the root at each reduced frequency, as the LEV remains attached in this region. Due to the presence of an outboard hairpin vortex over the moderately swept wing, there is a negative lift spike at a low reduced frequency. By contrast, the unswept wing LEV shows a positive lift further outboard, due to being attached to the leading edge near the wing tip at low reduced frequency.

Further research is planned to study intermediate reduced frequencies to ascertain if there is a more general relationship between reduced frequency and LEV structure. Aspect ratio effects are also to be considered as this has a major effect on LEV-TV interaction and the structure of the LEV. Finally, the effects of a thicker NACA0018 airfoil profile, particularly on LEV formation, are to be considered.

\begin{acknowledgments}
The authors gratefully acknowledge the support of the UK Engineering and Physical Sciences Research Council (EPSRC) through a DTA scholarship, grant EP/T517896/1. This work is supported by the US Air Force Office of Scientific Research under award number FA8655-21-1-7018, monitored by Dr. Douglas Smith. The work used the ARCHER2 UK National Supercomputing Service (https://www.archer2.ac.uk).
\end{acknowledgments}

\section*{Author Declarations}

\subsection*{Conflict of Interest}

The authors have no conflicts to disclose.

\subsection*{Author Contributions}

\textbf{Alex Cavanagh:} Conceptualization (equal); Data curation; Formal analysis (equal); Funding acquisition (supporting), Investigation; Methodology (equal); Software (lead); Validation (lead). \textbf{Chandan Bose:} Conceptualization (equal); Formal analysis (equal); Methodology (equal); Software (supporting); Supervision (supporting); Validation (supporting). \textbf{Kiran Ramesh:} Conceptualization (equal); Formal analysis (equal); Funding acquisition (lead); Methodology (equal); Software (supporting); Supervision (lead); Validation (supporting).

\section*{Data Availability Statement}

The data that support the findings of this study are available from the corresponding author upon reasonable request.

\bibliography{aipsamp}

\begin{thebibliography}{71}%
\makeatletter
\providecommand \@ifxundefined [1]{%
 \@ifx{#1\undefined}
}%
\providecommand \@ifnum [1]{%
 \ifnum #1\expandafter \@firstoftwo
 \else \expandafter \@secondoftwo
 \fi
}%
\providecommand \@ifx [1]{%
 \ifx #1\expandafter \@firstoftwo
 \else \expandafter \@secondoftwo
 \fi
}%
\providecommand \natexlab [1]{#1}%
\providecommand \enquote  [1]{``#1''}%
\providecommand \bibnamefont  [1]{#1}%
\providecommand \bibfnamefont [1]{#1}%
\providecommand \citenamefont [1]{#1}%
\providecommand \href@noop [0]{\@secondoftwo}%
\providecommand \href [0]{\begingroup \@sanitize@url \@href}%
\providecommand \@href[1]{\@@startlink{#1}\@@href}%
\providecommand \@@href[1]{\endgroup#1\@@endlink}%
\providecommand \@sanitize@url [0]{\catcode `\\12\catcode `\$12\catcode `\&12\catcode `\#12\catcode `\^12\catcode `\_12\catcode `\%12\relax}%
\providecommand \@@startlink[1]{}%
\providecommand \@@endlink[0]{}%
\providecommand \url  [0]{\begingroup\@sanitize@url \@url }%
\providecommand \@url [1]{\endgroup\@href {#1}{\urlprefix }}%
\providecommand \urlprefix  [0]{URL }%
\providecommand \Eprint [0]{\href }%
\providecommand \doibase [0]{http://dx.doi.org/}%
\providecommand \selectlanguage [0]{\@gobble}%
\providecommand \bibinfo  [0]{\@secondoftwo}%
\providecommand \bibfield  [0]{\@secondoftwo}%
\providecommand \translation [1]{[#1]}%
\providecommand \BibitemOpen [0]{}%
\providecommand \bibitemStop [0]{}%
\providecommand \bibitemNoStop [0]{.\EOS\space}%
\providecommand \EOS [0]{\spacefactor3000\relax}%
\providecommand \BibitemShut  [1]{\csname bibitem#1\endcsname}%
\let\auto@bib@innerbib\@empty
\bibitem [{\citenamefont {Ellington}\ \emph {et~al.}(1996)\citenamefont {Ellington}, \citenamefont {van~den Berg}, \citenamefont {Willmott},\ and\ \citenamefont {Thomas}}]{ellington_1996}%
  \BibitemOpen
  \bibfield  {author} {\bibinfo {author} {\bibfnamefont {C.~P.}\ \bibnamefont {Ellington}}, \bibinfo {author} {\bibfnamefont {C.}~\bibnamefont {van~den Berg}}, \bibinfo {author} {\bibfnamefont {A.~P.}\ \bibnamefont {Willmott}}, \ and\ \bibinfo {author} {\bibfnamefont {A.~L.~R.}\ \bibnamefont {Thomas}},\ }\bibfield  {title} {\enquote {\bibinfo {title} {Leading-edge vortices in insect flight},}\ }\href@noop {} {\bibfield  {journal} {\bibinfo  {journal} {Nature}\ }\textbf {\bibinfo {volume} {384}},\ \bibinfo {pages} {626--630} (\bibinfo {year} {1996})}\BibitemShut {NoStop}%
\bibitem [{\citenamefont {McCroskey}(1982)}]{mccroskey_1982}%
  \BibitemOpen
  \bibfield  {author} {\bibinfo {author} {\bibfnamefont {W.~J.}\ \bibnamefont {McCroskey}},\ }\bibfield  {title} {\enquote {\bibinfo {title} {Unsteady airfoils},}\ }\href@noop {} {\bibfield  {journal} {\bibinfo  {journal} {Annual Review of Fluid Mechanics}\ }\textbf {\bibinfo {volume} {14}},\ \bibinfo {pages} {285--311} (\bibinfo {year} {1982})}\BibitemShut {NoStop}%
\bibitem [{\citenamefont {{Anderson, Jr.}}(1991)}]{anderson_1991}%
  \BibitemOpen
  \bibfield  {author} {\bibinfo {author} {\bibfnamefont {J.~D.}\ \bibnamefont {{Anderson, Jr.}}},\ }\href@noop {} {\emph {\bibinfo {title} {Fundamentals of Aerodynamics}}},\ \bibinfo {edition} {2nd}\ ed.\ (\bibinfo  {publisher} {McGraw Hill, Inc.},\ \bibinfo {address} {New York},\ \bibinfo {year} {1991})\BibitemShut {NoStop}%
\bibitem [{\citenamefont {McCroskey}(1981)}]{mccroskey_1981}%
  \BibitemOpen
  \bibfield  {author} {\bibinfo {author} {\bibfnamefont {W.~J.}\ \bibnamefont {McCroskey}},\ }\href@noop {} {\enquote {\bibinfo {title} {The phenomenon of dynamic stall},}\ }\bibinfo {type} {Technical Memorandum}\ \bibinfo {number} {81264}\ (\bibinfo  {institution} {NASA},\ \bibinfo {year} {1981})\BibitemShut {NoStop}%
\bibitem [{\citenamefont {Eldredge}\ and\ \citenamefont {Jones}(2019)}]{eldredge&jones_2019}%
  \BibitemOpen
  \bibfield  {author} {\bibinfo {author} {\bibfnamefont {J.~D.}\ \bibnamefont {Eldredge}}\ and\ \bibinfo {author} {\bibfnamefont {A.~R.}\ \bibnamefont {Jones}},\ }\bibfield  {title} {\enquote {\bibinfo {title} {Leading-{Edge} {Vortices}: {Mechanics} and {Modeling}},}\ }\href@noop {} {\bibfield  {journal} {\bibinfo  {journal} {Annual Review of Fluid Mechanics}\ }\textbf {\bibinfo {volume} {51}},\ \bibinfo {pages} {75--104} (\bibinfo {year} {2019})}\BibitemShut {NoStop}%
\bibitem [{\citenamefont {Granlund}, \citenamefont {Ol},\ and\ \citenamefont {Bernal}(2013)}]{granlund_2013}%
  \BibitemOpen
  \bibfield  {author} {\bibinfo {author} {\bibfnamefont {K.~O.}\ \bibnamefont {Granlund}}, \bibinfo {author} {\bibfnamefont {M.~V.}\ \bibnamefont {Ol}}, \ and\ \bibinfo {author} {\bibfnamefont {L.~P.}\ \bibnamefont {Bernal}},\ }\bibfield  {title} {\enquote {\bibinfo {title} {Unsteady pitching flat plates},}\ }\href@noop {} {\bibfield  {journal} {\bibinfo  {journal} {Journal of Fluid Mechanics}\ }\textbf {\bibinfo {volume} {733}} (\bibinfo {year} {2013})}\BibitemShut {NoStop}%
\bibitem [{\citenamefont {Stevens}\ and\ \citenamefont {Babinsky}(2017)}]{stevens&babinsky_2017}%
  \BibitemOpen
  \bibfield  {author} {\bibinfo {author} {\bibfnamefont {P.~R. R.~J.}\ \bibnamefont {Stevens}}\ and\ \bibinfo {author} {\bibfnamefont {H.}~\bibnamefont {Babinsky}},\ }\bibfield  {title} {\enquote {\bibinfo {title} {Experiments to investigate lift production mechanisms on pitching flat plates},}\ }\href@noop {} {\bibfield  {journal} {\bibinfo  {journal} {Experiments in Fluids}\ }\textbf {\bibinfo {volume} {58}} (\bibinfo {year} {2017})}\BibitemShut {NoStop}%
\bibitem [{\citenamefont {Yilmaz}\ and\ \citenamefont {Rockwell}(2012)}]{yilmaz&rockwell_2012}%
  \BibitemOpen
  \bibfield  {author} {\bibinfo {author} {\bibfnamefont {T.~O.}\ \bibnamefont {Yilmaz}}\ and\ \bibinfo {author} {\bibfnamefont {D.}~\bibnamefont {Rockwell}},\ }\bibfield  {title} {\enquote {\bibinfo {title} {Flow structure on finite-span wings due to pitch-up motion},}\ }\href@noop {} {\bibfield  {journal} {\bibinfo  {journal} {Journal of Fluid Mechanics}\ }\textbf {\bibinfo {volume} {691}},\ \bibinfo {pages} {518--545} (\bibinfo {year} {2012})}\BibitemShut {NoStop}%
\bibitem [{\citenamefont {Eldredge}, \citenamefont {Wang},\ and\ \citenamefont {Ol}(2009)}]{eldredge_2009}%
  \BibitemOpen
  \bibfield  {author} {\bibinfo {author} {\bibfnamefont {J.~D.}\ \bibnamefont {Eldredge}}, \bibinfo {author} {\bibfnamefont {C.}~\bibnamefont {Wang}}, \ and\ \bibinfo {author} {\bibfnamefont {M.~V.}\ \bibnamefont {Ol}},\ }\bibfield  {title} {\enquote {\bibinfo {title} {A {Computational} {Study} of a {Canonical} {Pitch}-{Up}, {Pitch}-{Down} {Wing} {Maneuver}},}\ }in\ \href@noop {} {\emph {\bibinfo {booktitle} {39th AIAA Fluid Dynamics Conference}}},\ \bibinfo {series and number} {\bibinfo {series} {AIAA Meeting Paper}\ No.\ \bibinfo {number} {3687}}\ (\bibinfo {year} {2009})\BibitemShut {NoStop}%
\bibitem [{\citenamefont {Jantzen}\ \emph {et~al.}(2014)\citenamefont {Jantzen}, \citenamefont {Taira}, \citenamefont {Granlund},\ and\ \citenamefont {Ol}}]{jantzen_2014}%
  \BibitemOpen
  \bibfield  {author} {\bibinfo {author} {\bibfnamefont {R.~T.}\ \bibnamefont {Jantzen}}, \bibinfo {author} {\bibfnamefont {K.}~\bibnamefont {Taira}}, \bibinfo {author} {\bibfnamefont {K.~O.}\ \bibnamefont {Granlund}}, \ and\ \bibinfo {author} {\bibfnamefont {M.~V.}\ \bibnamefont {Ol}},\ }\bibfield  {title} {\enquote {\bibinfo {title} {Vortex dynamics around pitching plates},}\ }\href@noop {} {\bibfield  {journal} {\bibinfo  {journal} {Physics of Fluids}\ }\textbf {\bibinfo {volume} {26}} (\bibinfo {year} {2014})}\BibitemShut {NoStop}%
\bibitem [{\citenamefont {Visbal}(2017)}]{visbal_2017}%
  \BibitemOpen
  \bibfield  {author} {\bibinfo {author} {\bibfnamefont {M.~R.}\ \bibnamefont {Visbal}},\ }\bibfield  {title} {\enquote {\bibinfo {title} {Unsteady flow structure and loading of a pitching low-aspect-ratio wing},}\ }\href@noop {} {\bibfield  {journal} {\bibinfo  {journal} {Physical Review Fluids}\ }\textbf {\bibinfo {volume} {2}} (\bibinfo {year} {2017})}\BibitemShut {NoStop}%
\bibitem [{\citenamefont {Green}, \citenamefont {Rowley},\ and\ \citenamefont {Smits}(2011)}]{green_2011}%
  \BibitemOpen
  \bibfield  {author} {\bibinfo {author} {\bibfnamefont {M.~A.}\ \bibnamefont {Green}}, \bibinfo {author} {\bibfnamefont {C.~W.}\ \bibnamefont {Rowley}}, \ and\ \bibinfo {author} {\bibfnamefont {A.~J.}\ \bibnamefont {Smits}},\ }\bibfield  {title} {\enquote {\bibinfo {title} {The unsteady three-dimensional wake produced by a trapezoidal pitching panel},}\ }\href@noop {} {\bibfield  {journal} {\bibinfo  {journal} {Journal of Fluid Mechanics}\ }\textbf {\bibinfo {volume} {685}},\ \bibinfo {pages} {117--145} (\bibinfo {year} {2011})}\BibitemShut {NoStop}%
\bibitem [{\citenamefont {Hord}\ and\ \citenamefont {Lian}(2016)}]{hord&lian_2016}%
  \BibitemOpen
  \bibfield  {author} {\bibinfo {author} {\bibfnamefont {K.}~\bibnamefont {Hord}}\ and\ \bibinfo {author} {\bibfnamefont {Y.}~\bibnamefont {Lian}},\ }\bibfield  {title} {\enquote {\bibinfo {title} {Leading {Edge} {Vortex} {Circulation} {Development} on {Finite} {Aspect} {Ratio} {Pitch}-{Up} {Wings}},}\ }\href@noop {} {\bibfield  {journal} {\bibinfo  {journal} {AIAA Journal}\ }\textbf {\bibinfo {volume} {54}},\ \bibinfo {pages} {2755--2767} (\bibinfo {year} {2016})}\BibitemShut {NoStop}%
\bibitem [{\citenamefont {Yilmaz}, \citenamefont {Ol},\ and\ \citenamefont {Rockwell}(2010)}]{yilmaz_2010}%
  \BibitemOpen
  \bibfield  {author} {\bibinfo {author} {\bibfnamefont {T.}~\bibnamefont {Yilmaz}}, \bibinfo {author} {\bibfnamefont {M.}~\bibnamefont {Ol}}, \ and\ \bibinfo {author} {\bibfnamefont {D.}~\bibnamefont {Rockwell}},\ }\bibfield  {title} {\enquote {\bibinfo {title} {Scaling of flow separation on a pitching low aspect ratio plate},}\ }\href@noop {} {\bibfield  {journal} {\bibinfo  {journal} {Journal of Fluids and Structures}\ }\textbf {\bibinfo {volume} {26}},\ \bibinfo {pages} {1034--1041} (\bibinfo {year} {2010})}\BibitemShut {NoStop}%
\bibitem [{\citenamefont {Calderon}\ \emph {et~al.}(2013)\citenamefont {Calderon}, \citenamefont {Wang}, \citenamefont {Gursul},\ and\ \citenamefont {Visbal}}]{calderon_2013}%
  \BibitemOpen
  \bibfield  {author} {\bibinfo {author} {\bibfnamefont {D.~E.}\ \bibnamefont {Calderon}}, \bibinfo {author} {\bibfnamefont {Z.}~\bibnamefont {Wang}}, \bibinfo {author} {\bibfnamefont {I.}~\bibnamefont {Gursul}}, \ and\ \bibinfo {author} {\bibfnamefont {M.~R.}\ \bibnamefont {Visbal}},\ }\bibfield  {title} {\enquote {\bibinfo {title} {Volumetric measurements and simulations of the vortex structures generated by low aspect ratio plunging wings},}\ }\href@noop {} {\bibfield  {journal} {\bibinfo  {journal} {Physics of Fluids}\ }\textbf {\bibinfo {volume} {25}} (\bibinfo {year} {2013})}\BibitemShut {NoStop}%
\bibitem [{\citenamefont {Calderon}, \citenamefont {Wang},\ and\ \citenamefont {Gursul}(2013)}]{calderon_2013a}%
  \BibitemOpen
  \bibfield  {author} {\bibinfo {author} {\bibfnamefont {D.~E.}\ \bibnamefont {Calderon}}, \bibinfo {author} {\bibfnamefont {Z.}~\bibnamefont {Wang}}, \ and\ \bibinfo {author} {\bibfnamefont {I.}~\bibnamefont {Gursul}},\ }\bibfield  {title} {\enquote {\bibinfo {title} {Lift-{Enhancing} {Vortex} {Flows} {Generated} by {Plunging} {Rectangular} {Wings} with {Small} {Amplitude}},}\ }\href@noop {} {\bibfield  {journal} {\bibinfo  {journal} {AIAA Journal}\ }\textbf {\bibinfo {volume} {51}},\ \bibinfo {pages} {2953--2964} (\bibinfo {year} {2013})}\BibitemShut {NoStop}%
\bibitem [{\citenamefont {Calderon}\ \emph {et~al.}(2014)\citenamefont {Calderon}, \citenamefont {Cleaver}, \citenamefont {Gursul},\ and\ \citenamefont {Wang}}]{calderon_2014}%
  \BibitemOpen
  \bibfield  {author} {\bibinfo {author} {\bibfnamefont {D.~E.}\ \bibnamefont {Calderon}}, \bibinfo {author} {\bibfnamefont {D.~J.}\ \bibnamefont {Cleaver}}, \bibinfo {author} {\bibfnamefont {I.}~\bibnamefont {Gursul}}, \ and\ \bibinfo {author} {\bibfnamefont {Z.}~\bibnamefont {Wang}},\ }\bibfield  {title} {\enquote {\bibinfo {title} {On the absence of asymmetric wakes for periodically plunging finite wings},}\ }\href@noop {} {\bibfield  {journal} {\bibinfo  {journal} {Physics of Fluids}\ }\textbf {\bibinfo {volume} {26}} (\bibinfo {year} {2014})}\BibitemShut {NoStop}%
\bibitem [{\citenamefont {Fishman}, \citenamefont {Wolfinger},\ and\ \citenamefont {Rockwell}(2017)}]{fishman_2017}%
  \BibitemOpen
  \bibfield  {author} {\bibinfo {author} {\bibfnamefont {G.}~\bibnamefont {Fishman}}, \bibinfo {author} {\bibfnamefont {M.}~\bibnamefont {Wolfinger}}, \ and\ \bibinfo {author} {\bibfnamefont {D.}~\bibnamefont {Rockwell}},\ }\bibfield  {title} {\enquote {\bibinfo {title} {The structure of a trailing vortex from a perturbed wing},}\ }\href@noop {} {\bibfield  {journal} {\bibinfo  {journal} {Journal of Fluid Mechanics}\ }\textbf {\bibinfo {volume} {824}},\ \bibinfo {pages} {701--721} (\bibinfo {year} {2017})}\BibitemShut {NoStop}%
\bibitem [{\citenamefont {Visbal}, \citenamefont {Yilmaz},\ and\ \citenamefont {Rockwell}(2013)}]{visbal_2013}%
  \BibitemOpen
  \bibfield  {author} {\bibinfo {author} {\bibfnamefont {M.}~\bibnamefont {Visbal}}, \bibinfo {author} {\bibfnamefont {T.~O.}\ \bibnamefont {Yilmaz}}, \ and\ \bibinfo {author} {\bibfnamefont {D.}~\bibnamefont {Rockwell}},\ }\bibfield  {title} {\enquote {\bibinfo {title} {Three-dimensional vortex formation on a heaving low-aspect-ratio wing: {Computations} and experiments},}\ }\href@noop {} {\bibfield  {journal} {\bibinfo  {journal} {Journal of Fluids and Structures}\ }\textbf {\bibinfo {volume} {38}},\ \bibinfo {pages} {58--76} (\bibinfo {year} {2013})}\BibitemShut {NoStop}%
\bibitem [{\citenamefont {Yilmaz}\ and\ \citenamefont {Rockwell}(2010)}]{yilmaz&rockwell_2010}%
  \BibitemOpen
  \bibfield  {author} {\bibinfo {author} {\bibfnamefont {T.~O.}\ \bibnamefont {Yilmaz}}\ and\ \bibinfo {author} {\bibfnamefont {D.}~\bibnamefont {Rockwell}},\ }\bibfield  {title} {\enquote {\bibinfo {title} {Three-dimensional flow structure on a maneuvering wing},}\ }\href@noop {} {\bibfield  {journal} {\bibinfo  {journal} {Experiments in Fluids}\ }\textbf {\bibinfo {volume} {48}},\ \bibinfo {pages} {539--544} (\bibinfo {year} {2010})}\BibitemShut {NoStop}%
\bibitem [{\citenamefont {Beals}\ and\ \citenamefont {Jones}(2015)}]{beals&jones_2015}%
  \BibitemOpen
  \bibfield  {author} {\bibinfo {author} {\bibfnamefont {N.}~\bibnamefont {Beals}}\ and\ \bibinfo {author} {\bibfnamefont {A.~R.}\ \bibnamefont {Jones}},\ }\bibfield  {title} {\enquote {\bibinfo {title} {Lift {Production} by a {Passively} {Flexible} {Rotating} {Wing}},}\ }\href@noop {} {\bibfield  {journal} {\bibinfo  {journal} {AIAA Journal}\ }\textbf {\bibinfo {volume} {53}},\ \bibinfo {pages} {2995--3005} (\bibinfo {year} {2015})}\BibitemShut {NoStop}%
\bibitem [{\citenamefont {Carr}, \citenamefont {Chen},\ and\ \citenamefont {Ringuette}(2013)}]{carr_2013}%
  \BibitemOpen
  \bibfield  {author} {\bibinfo {author} {\bibfnamefont {Z.~R.}\ \bibnamefont {Carr}}, \bibinfo {author} {\bibfnamefont {C.}~\bibnamefont {Chen}}, \ and\ \bibinfo {author} {\bibfnamefont {M.~J.}\ \bibnamefont {Ringuette}},\ }\bibfield  {title} {\enquote {\bibinfo {title} {Finite-span rotating wings: three-dimensional vortex formation and variations with aspect ratio},}\ }\href@noop {} {\bibfield  {journal} {\bibinfo  {journal} {Experiments in Fluids}\ }\textbf {\bibinfo {volume} {54}} (\bibinfo {year} {2013})}\BibitemShut {NoStop}%
\bibitem [{\citenamefont {Carr}, \citenamefont {DeVoria},\ and\ \citenamefont {Ringuette}(2015)}]{carr_2015}%
  \BibitemOpen
  \bibfield  {author} {\bibinfo {author} {\bibfnamefont {Z.~R.}\ \bibnamefont {Carr}}, \bibinfo {author} {\bibfnamefont {A.~C.}\ \bibnamefont {DeVoria}}, \ and\ \bibinfo {author} {\bibfnamefont {M.~J.}\ \bibnamefont {Ringuette}},\ }\bibfield  {title} {\enquote {\bibinfo {title} {Aspect-ratio effects on rotating wings: circulation and forces},}\ }\href@noop {} {\bibfield  {journal} {\bibinfo  {journal} {Journal of Fluid Mechanics}\ }\textbf {\bibinfo {volume} {767}},\ \bibinfo {pages} {497--525} (\bibinfo {year} {2015})}\BibitemShut {NoStop}%
\bibitem [{\citenamefont {Medina}\ and\ \citenamefont {Jones}(2016)}]{medina&jones_2016}%
  \BibitemOpen
  \bibfield  {author} {\bibinfo {author} {\bibfnamefont {A.}~\bibnamefont {Medina}}\ and\ \bibinfo {author} {\bibfnamefont {A.~R.}\ \bibnamefont {Jones}},\ }\bibfield  {title} {\enquote {\bibinfo {title} {Leading-edge vortex burst on a low-aspect-ratio rotating flat plate},}\ }\href@noop {} {\bibfield  {journal} {\bibinfo  {journal} {Physical Review Fluids}\ }\textbf {\bibinfo {volume} {1}} (\bibinfo {year} {2016})}\BibitemShut {NoStop}%
\bibitem [{\citenamefont {Ozen}\ and\ \citenamefont {Rockwell}(2012)}]{ozen&rockwell_2012}%
  \BibitemOpen
  \bibfield  {author} {\bibinfo {author} {\bibfnamefont {C.~A.}\ \bibnamefont {Ozen}}\ and\ \bibinfo {author} {\bibfnamefont {D.}~\bibnamefont {Rockwell}},\ }\bibfield  {title} {\enquote {\bibinfo {title} {Three-dimensional vortex structure on a rotating wing},}\ }\href@noop {} {\bibfield  {journal} {\bibinfo  {journal} {Journal of Fluid Mechanics}\ }\textbf {\bibinfo {volume} {707}},\ \bibinfo {pages} {541--550} (\bibinfo {year} {2012})}\BibitemShut {NoStop}%
\bibitem [{\citenamefont {Venkata}\ and\ \citenamefont {Jones}(2013)}]{venkata&jones_2013}%
  \BibitemOpen
  \bibfield  {author} {\bibinfo {author} {\bibfnamefont {S.~K.}\ \bibnamefont {Venkata}}\ and\ \bibinfo {author} {\bibfnamefont {A.~R.}\ \bibnamefont {Jones}},\ }\bibfield  {title} {\enquote {\bibinfo {title} {Leading-{Edge} {Vortex} {Structure} over {Multiple} {Revolutions} of a {Rotating} {Wing}},}\ }\href@noop {} {\bibfield  {journal} {\bibinfo  {journal} {Journal of Aircraft}\ }\textbf {\bibinfo {volume} {50}},\ \bibinfo {pages} {1312--1316} (\bibinfo {year} {2013})}\BibitemShut {NoStop}%
\bibitem [{\citenamefont {DeVoria}\ and\ \citenamefont {Mohseni}(2017)}]{devoria&mohseni_2017}%
  \BibitemOpen
  \bibfield  {author} {\bibinfo {author} {\bibfnamefont {A.~C.}\ \bibnamefont {DeVoria}}\ and\ \bibinfo {author} {\bibfnamefont {K.}~\bibnamefont {Mohseni}},\ }\bibfield  {title} {\enquote {\bibinfo {title} {On the mechanism of high-incidence lift generation for steadily translating low-aspect-ratio wings},}\ }\href@noop {} {\bibfield  {journal} {\bibinfo  {journal} {Journal of Fluid Mechanics}\ }\textbf {\bibinfo {volume} {813}},\ \bibinfo {pages} {110--126} (\bibinfo {year} {2017})}\BibitemShut {NoStop}%
\bibitem [{\citenamefont {Mancini}\ \emph {et~al.}(2015)\citenamefont {Mancini}, \citenamefont {Manar}, \citenamefont {Granlund}, \citenamefont {Ol},\ and\ \citenamefont {Jones}}]{mancini_2015}%
  \BibitemOpen
  \bibfield  {author} {\bibinfo {author} {\bibfnamefont {P.}~\bibnamefont {Mancini}}, \bibinfo {author} {\bibfnamefont {F.}~\bibnamefont {Manar}}, \bibinfo {author} {\bibfnamefont {K.}~\bibnamefont {Granlund}}, \bibinfo {author} {\bibfnamefont {M.~V.}\ \bibnamefont {Ol}}, \ and\ \bibinfo {author} {\bibfnamefont {A.~R.}\ \bibnamefont {Jones}},\ }\bibfield  {title} {\enquote {\bibinfo {title} {Unsteady aerodynamic characteristics of a translating rigid wing at low {Reynolds} number},}\ }\href@noop {} {\bibfield  {journal} {\bibinfo  {journal} {Physics of Fluids}\ }\textbf {\bibinfo {volume} {27}} (\bibinfo {year} {2015})}\BibitemShut {NoStop}%
\bibitem [{\citenamefont {Mulleners}, \citenamefont {Mancini},\ and\ \citenamefont {Jones}(2017)}]{mulleners_2017}%
  \BibitemOpen
  \bibfield  {author} {\bibinfo {author} {\bibfnamefont {K.}~\bibnamefont {Mulleners}}, \bibinfo {author} {\bibfnamefont {P.}~\bibnamefont {Mancini}}, \ and\ \bibinfo {author} {\bibfnamefont {A.~R.}\ \bibnamefont {Jones}},\ }\bibfield  {title} {\enquote {\bibinfo {title} {Flow {Development} on a {Flat}-{Plate} {Wing} {Subjected} to a {Streamwise} {Acceleration}},}\ }\href@noop {} {\bibfield  {journal} {\bibinfo  {journal} {AIAA Journal}\ }\textbf {\bibinfo {volume} {55}},\ \bibinfo {pages} {2118--2122} (\bibinfo {year} {2017})}\BibitemShut {NoStop}%
\bibitem [{\citenamefont {Ol}\ \emph {et~al.}(2009)\citenamefont {Ol}, \citenamefont {Bernal}, \citenamefont {Kang},\ and\ \citenamefont {Shyy}}]{ol_2009}%
  \BibitemOpen
  \bibfield  {author} {\bibinfo {author} {\bibfnamefont {M.~V.}\ \bibnamefont {Ol}}, \bibinfo {author} {\bibfnamefont {L.}~\bibnamefont {Bernal}}, \bibinfo {author} {\bibfnamefont {C.-K.}\ \bibnamefont {Kang}}, \ and\ \bibinfo {author} {\bibfnamefont {W.}~\bibnamefont {Shyy}},\ }\bibfield  {title} {\enquote {\bibinfo {title} {Shallow and deep dynamic stall for flapping low {Reynolds} number airfoils},}\ }\href@noop {} {\bibfield  {journal} {\bibinfo  {journal} {Experiments in Fluids}\ }\textbf {\bibinfo {volume} {46}},\ \bibinfo {pages} {883--901} (\bibinfo {year} {2009})}\BibitemShut {NoStop}%
\bibitem [{\citenamefont {Manar}\ \emph {et~al.}(2016)\citenamefont {Manar}, \citenamefont {Mancini}, \citenamefont {Mayo},\ and\ \citenamefont {Jones}}]{manar_2016}%
  \BibitemOpen
  \bibfield  {author} {\bibinfo {author} {\bibfnamefont {F.}~\bibnamefont {Manar}}, \bibinfo {author} {\bibfnamefont {P.}~\bibnamefont {Mancini}}, \bibinfo {author} {\bibfnamefont {D.}~\bibnamefont {Mayo}}, \ and\ \bibinfo {author} {\bibfnamefont {A.~R.}\ \bibnamefont {Jones}},\ }\bibfield  {title} {\enquote {\bibinfo {title} {Comparison of {Rotating} and {Translating} {Wings}: {Force} {Production} and {Vortex} {Characteristics}},}\ }\href@noop {} {\bibfield  {journal} {\bibinfo  {journal} {AIAA Journal}\ }\textbf {\bibinfo {volume} {54}},\ \bibinfo {pages} {519--530} (\bibinfo {year} {2016})}\BibitemShut {NoStop}%
\bibitem [{\citenamefont {Percin}\ and\ \citenamefont {van Oudheusden}(2015)}]{percin&vanoudheusden_2015}%
  \BibitemOpen
  \bibfield  {author} {\bibinfo {author} {\bibfnamefont {M.}~\bibnamefont {Percin}}\ and\ \bibinfo {author} {\bibfnamefont {B.~W.}\ \bibnamefont {van Oudheusden}},\ }\bibfield  {title} {\enquote {\bibinfo {title} {Three-dimensional flow structures and unsteady forces on pitching and surging revolving flat plates},}\ }\href@noop {} {\bibfield  {journal} {\bibinfo  {journal} {Experiments in Fluids}\ }\textbf {\bibinfo {volume} {56}} (\bibinfo {year} {2015})}\BibitemShut {NoStop}%
\bibitem [{\citenamefont {Chellapurath}, \citenamefont {Noble},\ and\ \citenamefont {Sreejalekshmi}(2020)}]{chellapurath_2020}%
  \BibitemOpen
  \bibfield  {author} {\bibinfo {author} {\bibfnamefont {M.}~\bibnamefont {Chellapurath}}, \bibinfo {author} {\bibfnamefont {S.}~\bibnamefont {Noble}}, \ and\ \bibinfo {author} {\bibfnamefont {K.}~\bibnamefont {Sreejalekshmi}},\ }\bibfield  {title} {\enquote {\bibinfo {title} {Design and kinematic analysis of flapping wing mechanism for common swift inspired micro aerial vehicle},}\ }\href@noop {} {\bibfield  {journal} {\bibinfo  {journal} {Proceedings of the Institution of Mechanical Engineers, Part C: Journal of Mechanical Engineering Science}\ }\textbf {\bibinfo {volume} {0}},\ \bibinfo {pages} {1--11} (\bibinfo {year} {2020})}\BibitemShut {NoStop}%
\bibitem [{\citenamefont {Lentink}\ \emph {et~al.}(2007)\citenamefont {Lentink}, \citenamefont {Müller}, \citenamefont {Stamhuis}, \citenamefont {de~Kat}, \citenamefont {van Gestel}, \citenamefont {Veldhuis}, \citenamefont {Henningsson}, \citenamefont {Hedenström}, \citenamefont {Videler},\ and\ \citenamefont {van Leeuwen}}]{lentink_2007}%
  \BibitemOpen
  \bibfield  {author} {\bibinfo {author} {\bibfnamefont {D.}~\bibnamefont {Lentink}}, \bibinfo {author} {\bibfnamefont {U.~K.}\ \bibnamefont {Müller}}, \bibinfo {author} {\bibfnamefont {E.~J.}\ \bibnamefont {Stamhuis}}, \bibinfo {author} {\bibfnamefont {R.}~\bibnamefont {de~Kat}}, \bibinfo {author} {\bibfnamefont {W.}~\bibnamefont {van Gestel}}, \bibinfo {author} {\bibfnamefont {L.~L.~M.}\ \bibnamefont {Veldhuis}}, \bibinfo {author} {\bibfnamefont {P.}~\bibnamefont {Henningsson}}, \bibinfo {author} {\bibfnamefont {A.}~\bibnamefont {Hedenström}}, \bibinfo {author} {\bibfnamefont {J.~J.}\ \bibnamefont {Videler}}, \ and\ \bibinfo {author} {\bibfnamefont {J.~L.}\ \bibnamefont {van Leeuwen}},\ }\bibfield  {title} {\enquote {\bibinfo {title} {How swifts control their glide performance with morphing wings},}\ }\href@noop {} {\bibfield  {journal} {\bibinfo  {journal} {Nature}\ }\textbf {\bibinfo {volume} {446}},\ \bibinfo {pages} {1082--1085} (\bibinfo {year} {2007})}\BibitemShut {NoStop}%
\bibitem [{\citenamefont {Weiss}(2003)}]{weiss_2003}%
  \BibitemOpen
  \bibfield  {author} {\bibinfo {author} {\bibfnamefont {P.}~\bibnamefont {Weiss}},\ }\bibfield  {title} {\enquote {\bibinfo {title} {Wings of change: {Shape-shifting} aircraft may ply future skyways},}\ }\href@noop {} {\bibfield  {journal} {\bibinfo  {journal} {Science News}\ }\textbf {\bibinfo {volume} {164}},\ \bibinfo {pages} {359} (\bibinfo {year} {2003})}\BibitemShut {NoStop}%
\bibitem [{\citenamefont {Rayner}(1988)}]{rayner_1988}%
  \BibitemOpen
  \bibfield  {author} {\bibinfo {author} {\bibfnamefont {J.~M.~V.}\ \bibnamefont {Rayner}},\ }\enquote {\bibinfo {title} {Form and function in avian flight},}\ in\ \href@noop {} {\emph {\bibinfo {booktitle} {Current Ornithology}}},\ Vol.~\bibinfo {volume} {5},\ \bibinfo {editor} {edited by\ \bibinfo {editor} {\bibfnamefont {R.~F.}\ \bibnamefont {Johnston}}}\ (\bibinfo  {publisher} {Springer},\ \bibinfo {address} {Boston, MA},\ \bibinfo {year} {1988})\ Chap.~\bibinfo {chapter} {1}, pp.\ \bibinfo {pages} {1--66}\BibitemShut {NoStop}%
\bibitem [{\citenamefont {Azuma}(2006)}]{azuma_2006}%
  \BibitemOpen
  \bibfield  {author} {\bibinfo {author} {\bibfnamefont {A.}~\bibnamefont {Azuma}},\ }\href@noop {} {\emph {\bibinfo {title} {The Biokinetics of Flying and Swimming}}},\ \bibinfo {edition} {2nd}\ ed.\ (\bibinfo  {publisher} {American Institute of Aeronautics and Astronautics, Inc.},\ \bibinfo {address} {Reston, VA},\ \bibinfo {year} {2006})\BibitemShut {NoStop}%
\bibitem [{\citenamefont {Chiereghin}, \citenamefont {Cleaver},\ and\ \citenamefont {Gursul}(2017)}]{chiereghin_2017}%
  \BibitemOpen
  \bibfield  {author} {\bibinfo {author} {\bibfnamefont {N.}~\bibnamefont {Chiereghin}}, \bibinfo {author} {\bibfnamefont {D.}~\bibnamefont {Cleaver}}, \ and\ \bibinfo {author} {\bibfnamefont {I.}~\bibnamefont {Gursul}},\ }\bibfield  {title} {\enquote {\bibinfo {title} {Unsteady {Force} and {Flow} {Measurements} for {Plunging} {Finite} {Wings}},}\ }in\ \href@noop {} {\emph {\bibinfo {booktitle} {47th {AIAA} {Fluid} {Dynamics} {Conference}}}},\ \bibinfo {series and number} {{AIAA} {AVIATION} {Forum}}\ (\bibinfo  {publisher} {American Institute of Aeronautics and Astronautics},\ \bibinfo {year} {2017})\BibitemShut {NoStop}%
\bibitem [{\citenamefont {Chiereghin}\ \emph {et~al.}(2020)\citenamefont {Chiereghin}, \citenamefont {Bull}, \citenamefont {Cleaver},\ and\ \citenamefont {Gursul}}]{chiereghin_2020}%
  \BibitemOpen
  \bibfield  {author} {\bibinfo {author} {\bibfnamefont {N.}~\bibnamefont {Chiereghin}}, \bibinfo {author} {\bibfnamefont {S.}~\bibnamefont {Bull}}, \bibinfo {author} {\bibfnamefont {D.~J.}\ \bibnamefont {Cleaver}}, \ and\ \bibinfo {author} {\bibfnamefont {I.}~\bibnamefont {Gursul}},\ }\bibfield  {title} {\enquote {\bibinfo {title} {Three-dimensionality of leading-edge vortices on high aspect ratio plunging wings},}\ }\href@noop {} {\bibfield  {journal} {\bibinfo  {journal} {Physical Review Fluids}\ }\textbf {\bibinfo {volume} {5}},\ \bibinfo {pages} {064701} (\bibinfo {year} {2020})}\BibitemShut {NoStop}%
\bibitem [{\citenamefont {Son}\ \emph {et~al.}(2022)\citenamefont {Son}, \citenamefont {Gao}, \citenamefont {Gursul}, \citenamefont {Cantwell}, \citenamefont {Wang},\ and\ \citenamefont {Sherwin}}]{son_2022}%
  \BibitemOpen
  \bibfield  {author} {\bibinfo {author} {\bibfnamefont {O.}~\bibnamefont {Son}}, \bibinfo {author} {\bibfnamefont {A.-K.}\ \bibnamefont {Gao}}, \bibinfo {author} {\bibfnamefont {I.}~\bibnamefont {Gursul}}, \bibinfo {author} {\bibfnamefont {C.}~\bibnamefont {Cantwell}}, \bibinfo {author} {\bibfnamefont {Z.}~\bibnamefont {Wang}}, \ and\ \bibinfo {author} {\bibfnamefont {S.}~\bibnamefont {Sherwin}},\ }\bibfield  {title} {\enquote {\bibinfo {title} {Leading-edge vortex dynamics on plunging airfoils and wings},}\ }\href@noop {} {\bibfield  {journal} {\bibinfo  {journal} {Journal of Fluid Mechanics}\ }\textbf {\bibinfo {volume} {940}},\ \bibinfo {pages} {A28} (\bibinfo {year} {2022})}\BibitemShut {NoStop}%
\bibitem [{\citenamefont {Hammer}, \citenamefont {Garmann},\ and\ \citenamefont {Visbal}(2023)}]{hammer_2023}%
  \BibitemOpen
  \bibfield  {author} {\bibinfo {author} {\bibfnamefont {P.~R.}\ \bibnamefont {Hammer}}, \bibinfo {author} {\bibfnamefont {D.~J.}\ \bibnamefont {Garmann}}, \ and\ \bibinfo {author} {\bibfnamefont {M.~R.}\ \bibnamefont {Visbal}},\ }\bibfield  {title} {\enquote {\bibinfo {title} {Effect of {Aspect} {Ratio} on {Swept}-{Wing} {Dynamic} {Stall}},}\ }\href@noop {} {\bibfield  {journal} {\bibinfo  {journal} {AIAA Journal}\ }\textbf {\bibinfo {volume} {61}},\ \bibinfo {pages} {4367--4277} (\bibinfo {year} {2023})}\BibitemShut {NoStop}%
\bibitem [{\citenamefont {Garmann}\ and\ \citenamefont {Visbal}(2022)}]{garmann&visbal_2022}%
  \BibitemOpen
  \bibfield  {author} {\bibinfo {author} {\bibfnamefont {D.~J.}\ \bibnamefont {Garmann}}\ and\ \bibinfo {author} {\bibfnamefont {M.~R.}\ \bibnamefont {Visbal}},\ }\bibfield  {title} {\enquote {\bibinfo {title} {Control of {Dynamic} {Stall} on {Swept} {Finite} {Wings}},}\ }\href@noop {} {\bibfield  {journal} {\bibinfo  {journal} {AIAA Journal}\ ,\ \bibinfo {pages} {1--11}} (\bibinfo {year} {2022})}\BibitemShut {NoStop}%
\bibitem [{\citenamefont {Visbal}\ and\ \citenamefont {Garmann}(2019)}]{visbal&garmann_2019}%
  \BibitemOpen
  \bibfield  {author} {\bibinfo {author} {\bibfnamefont {M.~R.}\ \bibnamefont {Visbal}}\ and\ \bibinfo {author} {\bibfnamefont {D.~J.}\ \bibnamefont {Garmann}},\ }\bibfield  {title} {\enquote {\bibinfo {title} {Effect of {Sweep} on {Dynamic} {Stall} of a {Pitching} {Finite}-{Aspect}-{Ratio} {Wing}},}\ }\href@noop {} {\bibfield  {journal} {\bibinfo  {journal} {AIAA Journal}\ }\textbf {\bibinfo {volume} {57}},\ \bibinfo {pages} {3274--3289} (\bibinfo {year} {2019})}\BibitemShut {NoStop}%
\bibitem [{\citenamefont {Zhang}, \citenamefont {Hedrick},\ and\ \citenamefont {Mittal}(2015)}]{zhang_2015}%
  \BibitemOpen
  \bibfield  {author} {\bibinfo {author} {\bibfnamefont {C.}~\bibnamefont {Zhang}}, \bibinfo {author} {\bibfnamefont {T.~L.}\ \bibnamefont {Hedrick}}, \ and\ \bibinfo {author} {\bibfnamefont {R.}~\bibnamefont {Mittal}},\ }\bibfield  {title} {\enquote {\bibinfo {title} {Centripetal {Acceleration} {Reaction}: {An} {Effective} and {Robust} {Mechanism} for {Flapping} {Flight} in {Insects}},}\ }\href@noop {} {\bibfield  {journal} {\bibinfo  {journal} {PLoS ONE}\ }\textbf {\bibinfo {volume} {10}},\ \bibinfo {pages} {e0132093} (\bibinfo {year} {2015})}\BibitemShut {NoStop}%
\bibitem [{\citenamefont {W.~R.~Graham}\ and\ \citenamefont {Babinsky}(2017)}]{graham_2017}%
  \BibitemOpen
  \bibfield  {author} {\bibinfo {author} {\bibfnamefont {C.~W. P.~F.}\ \bibnamefont {W.~R.~Graham}}\ and\ \bibinfo {author} {\bibfnamefont {H.}~\bibnamefont {Babinsky}},\ }\bibfield  {title} {\enquote {\bibinfo {title} {An impulse-based approach to estimating forces in unsteady flow},}\ }\href@noop {} {\bibfield  {journal} {\bibinfo  {journal} {Journal of Fluid Mechanics}\ }\textbf {\bibinfo {volume} {815}},\ \bibinfo {pages} {60--76} (\bibinfo {year} {2017})}\BibitemShut {NoStop}%
\bibitem [{\citenamefont {J.~Li}(2018)}]{li&wu_2018}%
  \BibitemOpen
  \bibfield  {author} {\bibinfo {author} {\bibfnamefont {Z.~N.~W.}\ \bibnamefont {J.~Li}},\ }\bibfield  {title} {\enquote {\bibinfo {title} {Vortex force map method for viscous flows of general airfoils},}\ }\href@noop {} {\bibfield  {journal} {\bibinfo  {journal} {Journal of Fluid Mechanics}\ }\textbf {\bibinfo {volume} {836}},\ \bibinfo {pages} {145--166} (\bibinfo {year} {2018})}\BibitemShut {NoStop}%
\bibitem [{\citenamefont {Zhu}\ and\ \citenamefont {Breuer}(2023)}]{zhu&breuer_2023}%
  \BibitemOpen
  \bibfield  {author} {\bibinfo {author} {\bibfnamefont {Y.}~\bibnamefont {Zhu}}\ and\ \bibinfo {author} {\bibfnamefont {K.}~\bibnamefont {Breuer}},\ }\href@noop {} {\enquote {\bibinfo {title} {Flow-induced oscillations of pitching swept wings: {Stability} boundary, vortex dynamics and force partitioning},}\ } (\bibinfo {year} {2023})\BibitemShut {NoStop}%
\bibitem [{\citenamefont {Zhu}, \citenamefont {Mathai},\ and\ \citenamefont {Breuer}(2021)}]{zhu_2021}%
  \BibitemOpen
  \bibfield  {author} {\bibinfo {author} {\bibfnamefont {Y.}~\bibnamefont {Zhu}}, \bibinfo {author} {\bibfnamefont {V.}~\bibnamefont {Mathai}}, \ and\ \bibinfo {author} {\bibfnamefont {K.}~\bibnamefont {Breuer}},\ }\bibfield  {title} {\enquote {\bibinfo {title} {Nonlinear fluid damping of elastically mounted pitching wings in quiescent water},}\ }\href@noop {} {\bibfield  {journal} {\bibinfo  {journal} {Journal of Fluid Mechanics}\ }\textbf {\bibinfo {volume} {923}},\ \bibinfo {pages} {R2} (\bibinfo {year} {2021})}\BibitemShut {NoStop}%
\bibitem [{\citenamefont {Menon}, \citenamefont {Kumar},\ and\ \citenamefont {Mittal}(2022)}]{menon_2022}%
  \BibitemOpen
  \bibfield  {author} {\bibinfo {author} {\bibfnamefont {K.}~\bibnamefont {Menon}}, \bibinfo {author} {\bibfnamefont {S.}~\bibnamefont {Kumar}}, \ and\ \bibinfo {author} {\bibfnamefont {R.}~\bibnamefont {Mittal}},\ }\bibfield  {title} {\enquote {\bibinfo {title} {Contribution of spanwise and cross-span vortices to the lift generation of low-aspect-ratio wings: {Insights} from force partitioning},}\ }\href@noop {} {\bibfield  {journal} {\bibinfo  {journal} {Physical Review Fluids}\ }\textbf {\bibinfo {volume} {7}},\ \bibinfo {pages} {114102} (\bibinfo {year} {2022})}\BibitemShut {NoStop}%
\bibitem [{\citenamefont {Zhang}\ and\ \citenamefont {Taira}(2022)}]{zhang&taira_2022}%
  \BibitemOpen
  \bibfield  {author} {\bibinfo {author} {\bibfnamefont {K.}~\bibnamefont {Zhang}}\ and\ \bibinfo {author} {\bibfnamefont {K.}~\bibnamefont {Taira}},\ }\bibfield  {title} {\enquote {\bibinfo {title} {Laminar vortex dynamics around forward-swept wings},}\ }\href@noop {} {\bibfield  {journal} {\bibinfo  {journal} {Phys. Rev. Fluids}\ }\textbf {\bibinfo {volume} {7}},\ \bibinfo {pages} {024704} (\bibinfo {year} {2022})}\BibitemShut {NoStop}%
\bibitem [{\citenamefont {Li}, \citenamefont {Zhao},\ and\ \citenamefont {Graham}(2020)}]{li_2020}%
  \BibitemOpen
  \bibfield  {author} {\bibinfo {author} {\bibfnamefont {J.}~\bibnamefont {Li}}, \bibinfo {author} {\bibfnamefont {X.}~\bibnamefont {Zhao}}, \ and\ \bibinfo {author} {\bibfnamefont {M.}~\bibnamefont {Graham}},\ }\bibfield  {title} {\enquote {\bibinfo {title} {Vortex force maps for three-dimensional unsteady flows with application to a delta wing},}\ }\href@noop {} {\bibfield  {journal} {\bibinfo  {journal} {Journal of Fluid Mechanics}\ }\textbf {\bibinfo {volume} {900}},\ \bibinfo {pages} {A36} (\bibinfo {year} {2020})}\BibitemShut {NoStop}%
\bibitem [{\citenamefont {Ramesh}\ \emph {et~al.}(2014)\citenamefont {Ramesh}, \citenamefont {Gopalarathnam}, \citenamefont {Granlund}, \citenamefont {Ol},\ and\ \citenamefont {Edwards}}]{ramesh_2014}%
  \BibitemOpen
  \bibfield  {author} {\bibinfo {author} {\bibfnamefont {K.}~\bibnamefont {Ramesh}}, \bibinfo {author} {\bibfnamefont {A.}~\bibnamefont {Gopalarathnam}}, \bibinfo {author} {\bibfnamefont {K.}~\bibnamefont {Granlund}}, \bibinfo {author} {\bibfnamefont {M.~V.}\ \bibnamefont {Ol}}, \ and\ \bibinfo {author} {\bibfnamefont {J.~R.}\ \bibnamefont {Edwards}},\ }\bibfield  {title} {\enquote {\bibinfo {title} {Discrete-vortex method with novel shedding criterion for unsteady aerofoil flows with intermittent leading-edge vortex shedding},}\ }\href@noop {} {\bibfield  {journal} {\bibinfo  {journal} {Journal of Fluid Mechanics}\ }\textbf {\bibinfo {volume} {751}},\ \bibinfo {pages} {500--538} (\bibinfo {year} {2014})}\BibitemShut {NoStop}%
\bibitem [{\citenamefont {Maxworthy}(2007)}]{maxworthy_2007}%
  \BibitemOpen
  \bibfield  {author} {\bibinfo {author} {\bibfnamefont {T.}~\bibnamefont {Maxworthy}},\ }\bibfield  {title} {\enquote {\bibinfo {title} {The formation and maintenance of a leading-edge vortex during the forward motion of an animal wing},}\ }\href@noop {} {\bibfield  {journal} {\bibinfo  {journal} {Journal of Fluid Mechanics}\ }\textbf {\bibinfo {volume} {587}},\ \bibinfo {pages} {471--475} (\bibinfo {year} {2007})}\BibitemShut {NoStop}%
\bibitem [{\citenamefont {Harbig}, \citenamefont {Sheridan},\ and\ \citenamefont {Thompson}(2014)}]{harbig_2014}%
  \BibitemOpen
  \bibfield  {author} {\bibinfo {author} {\bibfnamefont {R.~R.}\ \bibnamefont {Harbig}}, \bibinfo {author} {\bibfnamefont {J.}~\bibnamefont {Sheridan}}, \ and\ \bibinfo {author} {\bibfnamefont {M.~C.}\ \bibnamefont {Thompson}},\ }\bibfield  {title} {\enquote {\bibinfo {title} {The role of advance ratio and aspect ratio in determining leading-edge vortex stability for flapping flight},}\ }\href@noop {} {\bibfield  {journal} {\bibinfo  {journal} {Journal of Fluid Mechanics}\ }\textbf {\bibinfo {volume} {751}},\ \bibinfo {pages} {71--105} (\bibinfo {year} {2014})}\BibitemShut {NoStop}%
\bibitem [{\citenamefont {Wojcik}\ and\ \citenamefont {Buchholz}(2014)}]{wojcik&buchholz_2014}%
  \BibitemOpen
  \bibfield  {author} {\bibinfo {author} {\bibfnamefont {C.~J.}\ \bibnamefont {Wojcik}}\ and\ \bibinfo {author} {\bibfnamefont {J.~H.~J.}\ \bibnamefont {Buchholz}},\ }\bibfield  {title} {\enquote {\bibinfo {title} {Vorticity transport in the leading-edge vortex on a rotating blade},}\ }\href@noop {} {\bibfield  {journal} {\bibinfo  {journal} {Journal of Fluid Mechanics}\ }\textbf {\bibinfo {volume} {743}},\ \bibinfo {pages} {249--261} (\bibinfo {year} {2014})}\BibitemShut {NoStop}%
\bibitem [{\citenamefont {Wong}\ and\ \citenamefont {Rival}(2015)}]{wong&rival_2015}%
  \BibitemOpen
  \bibfield  {author} {\bibinfo {author} {\bibfnamefont {J.}~\bibnamefont {Wong}}\ and\ \bibinfo {author} {\bibfnamefont {D.}~\bibnamefont {Rival}},\ }\bibfield  {title} {\enquote {\bibinfo {title} {Determining the relative stability of leading-edge vortices on nominally two-dimensional flapping profiles},}\ }\href@noop {} {\bibfield  {journal} {\bibinfo  {journal} {Journal of Fluid Mechanics}\ }\textbf {\bibinfo {volume} {766}},\ \bibinfo {pages} {611--625} (\bibinfo {year} {2015})}\BibitemShut {NoStop}%
\bibitem [{\citenamefont {Limacher}, \citenamefont {Morton},\ and\ \citenamefont {Wood}(2016)}]{limacher_2016}%
  \BibitemOpen
  \bibfield  {author} {\bibinfo {author} {\bibfnamefont {E.}~\bibnamefont {Limacher}}, \bibinfo {author} {\bibfnamefont {C.}~\bibnamefont {Morton}}, \ and\ \bibinfo {author} {\bibfnamefont {D.}~\bibnamefont {Wood}},\ }\bibfield  {title} {\enquote {\bibinfo {title} {On the trajectory of leading-edge vortices under the influence of {Coriolis} acceleration},}\ }\href@noop {} {\bibfield  {journal} {\bibinfo  {journal} {Journal of Fluid Mechanics}\ }\textbf {\bibinfo {volume} {800}},\ \bibinfo {pages} {R1} (\bibinfo {year} {2016})}\BibitemShut {NoStop}%
\bibitem [{\citenamefont {Hirato}\ \emph {et~al.}(2019)\citenamefont {Hirato}, \citenamefont {Shen}, \citenamefont {Gopalarathnam},\ and\ \citenamefont {Edwards}}]{hirato_2019}%
  \BibitemOpen
  \bibfield  {author} {\bibinfo {author} {\bibfnamefont {Y.}~\bibnamefont {Hirato}}, \bibinfo {author} {\bibfnamefont {M.}~\bibnamefont {Shen}}, \bibinfo {author} {\bibfnamefont {A.}~\bibnamefont {Gopalarathnam}}, \ and\ \bibinfo {author} {\bibfnamefont {J.~R.}\ \bibnamefont {Edwards}},\ }\bibfield  {title} {\enquote {\bibinfo {title} {Vortex-{Sheet} {Representation} of {Leading}-{Edge} {Vortex} {Shedding} from {Finite} {Wings}},}\ }\href@noop {} {\bibfield  {journal} {\bibinfo  {journal} {Journal of Aircraft}\ }\textbf {\bibinfo {volume} {56}},\ \bibinfo {pages} {1626--1640} (\bibinfo {year} {2019})}\BibitemShut {NoStop}%
\bibitem [{\citenamefont {Hirato}\ \emph {et~al.}(2021)\citenamefont {Hirato}, \citenamefont {Shen}, \citenamefont {Gopalarathnam},\ and\ \citenamefont {Edwards}}]{hirato_2021}%
  \BibitemOpen
  \bibfield  {author} {\bibinfo {author} {\bibfnamefont {Y.}~\bibnamefont {Hirato}}, \bibinfo {author} {\bibfnamefont {M.}~\bibnamefont {Shen}}, \bibinfo {author} {\bibfnamefont {A.}~\bibnamefont {Gopalarathnam}}, \ and\ \bibinfo {author} {\bibfnamefont {J.~R.}\ \bibnamefont {Edwards}},\ }\bibfield  {title} {\enquote {\bibinfo {title} {Flow criticality governs leading-edge-vortex initiation on finite wings in unsteady flow},}\ }\href@noop {} {\bibfield  {journal} {\bibinfo  {journal} {Journal of Fluid Mechanics}\ }\textbf {\bibinfo {volume} {910}},\ \bibinfo {pages} {A1} (\bibinfo {year} {2021})}\BibitemShut {NoStop}%
\bibitem [{\citenamefont {Aggarwal}(2013)}]{aggarwal_2013}%
  \BibitemOpen
  \bibfield  {author} {\bibinfo {author} {\bibfnamefont {S.}~\bibnamefont {Aggarwal}},\ }\emph {\bibinfo {title} {An Inviscid Numerical Method for Unsteady Flows over Airfoils and Wings to Predict the Onset of Leading Edge Vortex Formation}},\ \href@noop {} {\bibinfo {type} {Master of science}},\ \bibinfo  {school} {Graduate Faculty of North Carolina State University} (\bibinfo {year} {2013})\BibitemShut {NoStop}%
\bibitem [{\citenamefont {Ramesh}(2020)}]{ramesh_2020}%
  \BibitemOpen
  \bibfield  {author} {\bibinfo {author} {\bibfnamefont {K.}~\bibnamefont {Ramesh}},\ }\bibfield  {title} {\enquote {\bibinfo {title} {On the leading-edge suction and stagnation-point location in unsteady flows past thin aerofoils},}\ }\href@noop {} {\bibfield  {journal} {\bibinfo  {journal} {Journal of Fluid Mechanics}\ }\textbf {\bibinfo {volume} {886}},\ \bibinfo {pages} {A13} (\bibinfo {year} {2020})}\BibitemShut {NoStop}%
\bibitem [{\citenamefont {Martínez}\ \emph {et~al.}(2022)\citenamefont {Martínez}, \citenamefont {He}, \citenamefont {Mulleners},\ and\ \citenamefont {Ramesh}}]{martinez_2022}%
  \BibitemOpen
  \bibfield  {author} {\bibinfo {author} {\bibfnamefont {A.}~\bibnamefont {Martínez}}, \bibinfo {author} {\bibfnamefont {G.}~\bibnamefont {He}}, \bibinfo {author} {\bibfnamefont {K.}~\bibnamefont {Mulleners}}, \ and\ \bibinfo {author} {\bibfnamefont {K.~K.}\ \bibnamefont {Ramesh}},\ }\bibfield  {title} {\enquote {\bibinfo {title} {Modulation of the leading-edge vortex shedding rate in discrete-vortex methods},}\ }in\ \href@noop {} {\emph {\bibinfo {booktitle} {AIAA SCITECH 2022 Forum}}}\ (\bibinfo {year} {2022})\BibitemShut {NoStop}%
\bibitem [{\citenamefont {Fage}\ and\ \citenamefont {Johansen}(1927)}]{fage&johansen_1927}%
  \BibitemOpen
  \bibfield  {author} {\bibinfo {author} {\bibfnamefont {A.}~\bibnamefont {Fage}}\ and\ \bibinfo {author} {\bibfnamefont {F.~C.}\ \bibnamefont {Johansen}},\ }\bibfield  {title} {\enquote {\bibinfo {title} {On the flow of air behind an inclined flat plate of infinite span},}\ }\href@noop {} {\bibfield  {journal} {\bibinfo  {journal} {Proceedings of the Royal Society of London. Series A, Containing Papers of a Mathematical and Physical Character}\ }\textbf {\bibinfo {volume} {116}},\ \bibinfo {pages} {170--197} (\bibinfo {year} {1927})}\BibitemShut {NoStop}%
\bibitem [{\citenamefont {Menon}\ and\ \citenamefont {Mittal}(2021{\natexlab{a}})}]{menon&mittal_2021}%
  \BibitemOpen
  \bibfield  {author} {\bibinfo {author} {\bibfnamefont {K.}~\bibnamefont {Menon}}\ and\ \bibinfo {author} {\bibfnamefont {R.}~\bibnamefont {Mittal}},\ }\bibfield  {title} {\enquote {\bibinfo {title} {On the initiation and sustenance of flow-induced vibration of cylinders: insights from force partitioning},}\ }\href@noop {} {\bibfield  {journal} {\bibinfo  {journal} {Journal of Fluid Mechanics}\ }\textbf {\bibinfo {volume} {907}},\ \bibinfo {pages} {A37} (\bibinfo {year} {2021}{\natexlab{a}})}\BibitemShut {NoStop}%
\bibitem [{\citenamefont {Menon}\ and\ \citenamefont {Mittal}(2021{\natexlab{b}})}]{menon&mittal_2021b}%
  \BibitemOpen
  \bibfield  {author} {\bibinfo {author} {\bibfnamefont {K.}~\bibnamefont {Menon}}\ and\ \bibinfo {author} {\bibfnamefont {R.}~\bibnamefont {Mittal}},\ }\bibfield  {title} {\enquote {\bibinfo {title} {Quantitative analysis of the kinematics and induced aerodynamic loading of individual vortices in vortex-dominated flows: {A} computation and data-driven approach},}\ }\href@noop {} {\bibfield  {journal} {\bibinfo  {journal} {Journal of Computational Physics}\ }\textbf {\bibinfo {volume} {443}},\ \bibinfo {pages} {110515} (\bibinfo {year} {2021}{\natexlab{b}})}\BibitemShut {NoStop}%
\bibitem [{\citenamefont {Jeong}\ and\ \citenamefont {Hussain}(1995)}]{jeong&hussain_1995}%
  \BibitemOpen
  \bibfield  {author} {\bibinfo {author} {\bibfnamefont {J.}~\bibnamefont {Jeong}}\ and\ \bibinfo {author} {\bibfnamefont {F.}~\bibnamefont {Hussain}},\ }\bibfield  {title} {\enquote {\bibinfo {title} {On the identification of a vortex},}\ }\href@noop {} {\bibfield  {journal} {\bibinfo  {journal} {Journal of Fluid Mechanics}\ }\textbf {\bibinfo {volume} {285}},\ \bibinfo {pages} {69--94} (\bibinfo {year} {1995})}\BibitemShut {NoStop}%
\bibitem [{\citenamefont {Lentink}\ and\ \citenamefont {Dickinson}(2009)}]{lentink&dickinson_2009}%
  \BibitemOpen
  \bibfield  {author} {\bibinfo {author} {\bibfnamefont {D.}~\bibnamefont {Lentink}}\ and\ \bibinfo {author} {\bibfnamefont {M.~H.}\ \bibnamefont {Dickinson}},\ }\bibfield  {title} {\enquote {\bibinfo {title} {Rotational accelerations stabilize leading edge vortices on revolving fly wings},}\ }\href@noop {} {\bibfield  {journal} {\bibinfo  {journal} {Journal of Experimental Biology}\ }\textbf {\bibinfo {volume} {212}},\ \bibinfo {pages} {2705--2719} (\bibinfo {year} {2009})}\BibitemShut {NoStop}%
\bibitem [{\citenamefont {McGowan}\ \emph {et~al.}(2011)\citenamefont {McGowan}, \citenamefont {Granlund}, \citenamefont {Ol}, \citenamefont {Gopalarathnam},\ and\ \citenamefont {Edwards}}]{mcgowan_2011}%
  \BibitemOpen
  \bibfield  {author} {\bibinfo {author} {\bibfnamefont {G.~Z.}\ \bibnamefont {McGowan}}, \bibinfo {author} {\bibfnamefont {K.}~\bibnamefont {Granlund}}, \bibinfo {author} {\bibfnamefont {M.~V.}\ \bibnamefont {Ol}}, \bibinfo {author} {\bibfnamefont {A.}~\bibnamefont {Gopalarathnam}}, \ and\ \bibinfo {author} {\bibfnamefont {J.~R.}\ \bibnamefont {Edwards}},\ }\bibfield  {title} {\enquote {\bibinfo {title} {Investigations of {Lift}-{Based} {Pitch}-{Plunge} {Equivalence} for {Airfoils} at {Low} {Reynolds} {Numbers}},}\ }\href@noop {} {\bibfield  {journal} {\bibinfo  {journal} {AIAA Journal}\ }\textbf {\bibinfo {volume} {49}},\ \bibinfo {pages} {1511--1524} (\bibinfo {year} {2011})}\BibitemShut {NoStop}%
\bibitem [{\citenamefont {Shur}\ \emph {et~al.}(2008)\citenamefont {Shur}, \citenamefont {Spalart}, \citenamefont {Strelets},\ and\ \citenamefont {Travin}}]{shur_2008}%
  \BibitemOpen
  \bibfield  {author} {\bibinfo {author} {\bibfnamefont {M.~L.}\ \bibnamefont {Shur}}, \bibinfo {author} {\bibfnamefont {P.~R.}\ \bibnamefont {Spalart}}, \bibinfo {author} {\bibfnamefont {M.~K.}\ \bibnamefont {Strelets}}, \ and\ \bibinfo {author} {\bibfnamefont {A.~K.}\ \bibnamefont {Travin}},\ }\bibfield  {title} {\enquote {\bibinfo {title} {A hybrid {RANS}-{LES} approach with delayed-{DES} and wall-modelled {LES} capabilities},}\ }\href@noop {} {\bibfield  {journal} {\bibinfo  {journal} {International Journal of Heat and Fluid Flow}\ }\textbf {\bibinfo {volume} {29}},\ \bibinfo {pages} {1638--1649} (\bibinfo {year} {2008})}\BibitemShut {NoStop}%
\bibitem [{\citenamefont {Spalart}(2001)}]{spalart_2001}%
  \BibitemOpen
  \bibfield  {author} {\bibinfo {author} {\bibfnamefont {P.~R.}\ \bibnamefont {Spalart}},\ }\href@noop {} {\enquote {\bibinfo {title} {Young-person's guide to detached-eddy simulation grids},}\ }\bibinfo {type} {Contractor Report}\ \bibinfo {number} {211032}\ (\bibinfo  {institution} {NASA},\ \bibinfo {year} {2001})\BibitemShut {NoStop}%
\bibitem [{\citenamefont {Bird}\ \emph {et~al.}(2022)\citenamefont {Bird}, \citenamefont {Ramesh}, \citenamefont {Ōtomo},\ and\ \citenamefont {Maria~Viola}}]{bird_2022}%
  \BibitemOpen
  \bibfield  {author} {\bibinfo {author} {\bibfnamefont {H.~J.~A.}\ \bibnamefont {Bird}}, \bibinfo {author} {\bibfnamefont {K.}~\bibnamefont {Ramesh}}, \bibinfo {author} {\bibfnamefont {S.}~\bibnamefont {Ōtomo}}, \ and\ \bibinfo {author} {\bibfnamefont {I.}~\bibnamefont {Maria~Viola}},\ }\bibfield  {title} {\enquote {\bibinfo {title} {Usefulness of {Inviscid} {Linear} {Unsteady} {Lifting}-{Line} {Theory} for {Viscous} {Large}-{Amplitude} {Problems}},}\ }\href@noop {} {\bibfield  {journal} {\bibinfo  {journal} {AIAA Journal}\ }\textbf {\bibinfo {volume} {60}},\ \bibinfo {pages} {598--609} (\bibinfo {year} {2022})}\BibitemShut {NoStop}%
\end{thebibliography}%

\end{document}